\documentclass[longauth]{aa} % for the long lists of affiliations 
\usepackage{graphicx}
\usepackage[flushleft]{threeparttable}
\usepackage{txfonts}
\usepackage{hyperref}
\def\bpic{$\beta$ Pictoris}
\def\bpicb{$\beta$ Pictoris b}
\def\bpicc{$\beta$ Pictoris c}
\def\dsct{$\delta$~Scuti\,}

\def\Msun{$M_{\odot}$}
\def\MJ{$M_{J}$}

\def\Rsun{$R_{\odot}$}
\def\Msun{$M_{\odot}$}

\def\cd{\,d$^{\rm -1}$}

\def\kms{$\mathrm{km\,s}^{-1}$}

\graphicspath{ {./final_figs/} }

\begin{document} 

\title{The $\beta$ Pictoris b Hill sphere transit campaign}
\subtitle{II. Searching for the signatures of the \bpic{} exoplanets \\through time delay analysis of the $\delta$ Scuti pulsations}
\titlerunning{Time delay analysis of \bpic}

\author{Sebastian Zieba\inst{1,2,3}
\and
Konstanze Zwintz\inst{3}
\and
Matthew Kenworthy\inst{2}
\and
Daniel Hey\inst{4}
\and
Simon J. Murphy\inst{5}
\and
Rainer Kuschnig\inst{6}
\and
Lyu Abe\inst{7}
\and
Abdelkrim Agabi\inst{7}
\and
Djamel Mekarnia\inst{7}
\and
Tristan Guillot\inst{7}
\and
Fran\c{c}ois-Xavier Schmider\inst{7}
\and
Philippe Stee\inst{7}
\and
Yuri De Pra\inst{8}
\and
Marco Buttu\inst{9,10}
\and
Nicolas Crouzet\inst{2}
\and
Samuel Mellon\inst{11}
\and
Jeb Bailey III\inst{12}
\and
Remko Stuik\inst{2}
\and
Patrick Dorval\inst{13}
\and
Geert-Jan J. Talens\inst{14}
\and
Steven Crawford\inst{15}
\and
Eric Mamajek\inst{11,16}
\and
Iva Laginja\inst{17}
\and
Michael Ireland\inst{18}
\and
Blaine Lomberg\inst{19,20}
\and
Rudi Kuhn\inst{19}
\and
Ignas Snellen\inst{2}
\and
Paul Kalas\inst{22,23,24}
\and
Jason J. Wang\inst{25}
\and
Kevin B. Stevenson\inst{26}
\and
Ernst de Mooij\inst{27,28}
\and
Anne-Marie Lagrange\inst{29,30,31}
\and
Sylvestre Lacour\inst{17}
\and
Mathias Nowak\inst{31}
\and
Paul A. Str\o{}m\inst{32,33}
\and
Zhang Hui\inst{34}
\and
Lifan Wang\inst{35}
}

\institute{Max-Planck-Institut f\"ur Astronomie, K\"onigstuhl 17, D-69117 Heidelberg, Germany
 \and
Leiden Observatory, Leiden University, Postbus 9513, 2300 RA Leiden, The Netherlands 
 \and
Institut f\"ur Astro- und Teilchenphysik, Universit\"at Innsbruck, Technikerstra{\ss}e 25, A-6020 Innsbruck
 \and
 Institute for Astronomy, University of Hawai‘i, Honolulu, HI 96822, USA
 \and
 Centre for Astrophysics, University of Southern Queensland, Toowoomba, QLD 4350, Australia
 \and
Institut für Physik, Karl-Franzens Universit\"at Graz, Universit\"atsplatz 5/II, NAWI Graz, 8010 Graz, Austria 
\and
Universit\'{e} C\^{o}te d'Azur, Observatoire de la C\^{o}te d'Azur, CNRS, Laboratoire Lagrange, France
 \and
 Italian National Agency for New Technologies, Energy and Sustainable Economic Development (ENEA), via Anguillarese 301, Rome, Italy.
 \and
Institut polaire français Paul Émile Victor and Programma Nazionale di Ricerche in Antartide, Concordia Station, Antarctica
 \and
Istituto Nazionale di Astrofisica, Osservatorio Astronomico di Cagliari, Italy
 \and
Department of Physics \& Astronomy, University of Rochester, Rochester, NY 14627, USA
 \and
Department of Physics, University of California at Santa Barbara, Santa Barbara, CA 93106, USA
 \and
Department of Physics and Astronomy, Astronomy and Space Physics, Uppsala University, 751 20 Uppsala, Sweden
 \and
Institut de Recherche sur les Exoplan\`{e}tes, D\'{e}partement de Physique, Universit\'{e} de Montr\'{e}al, Montr\'{e}al, QC H3C 3J7, Canada
 \and
unaffiliated 
 \and
Jet Propulsion Laboratory, California Institute of Technology, 4800 Oak Grove Drive, M/S321-100, Pasadena, CA 91109, USA
 \and
LESIA, Observatoire de Paris, Universit\'{e} PSL, CNRS, Sorbonne Universit\'{e}, Universit\'{e} de Paris, 5 place Jules Janssen, 92195 Meudon, France
 \and
Research School of Astronomy and Astrophysics, Australian National University, Canberra, ACT 2611, Australia
 \and
 South African Astronomical Observatory, Observatory Rd, Observatory Cape Town, 7700 Cape Town, South Africa
 \and
Department of Astronomy, University of Cape Town, Rondebosch, 7700 Cape Town, South Africa
 \and
Astronomy Department, University of California, Berkeley, CA 94720, USA
 \and
SETI Institute, Carl Sagan Center, 189 Bernardo Ave.,  Mountain View CA 94043, USA
 \and
Institute of Astrophysics, FORTH, GR-71110 Heraklion, Greece
 \and
Center for Interdisciplinary Exploration and Research in Astrophysics (CIERA) and Department of Physics and Astronomy, Northwestern University, Evanston, IL 60208, USA
 \and
 Johns Hopkins APL, Laurel, MD, USA
 \and
Astrophysics Research Centre, Queen’s University Belfast, Belfast BT7 1NN, UK
 \and
School of Physical Sciences and Centre for Astrophysics \& Relativity, Dublin City University, Glasnevin, Dublin 9, Ireland
 \and
IPAG, Univ. Grenoble Alpes, CNRS, IPAG, F-38000 Grenoble, France
 \and
IMCCE - Observatoire de Paris, 77 Avenue Denfert-Rochereau, F-75014 PARIS
 \and
Institut d’Astrophysique de Paris, UMR7095 CNRS, Universit\'{e} Pierre \& Marie Curie, 98 bis boulevard Arago, 75014 Paris, France
 \and
Institute of Astronomy, Madingley Road, Cambridge CB3 0HA, UK
 \and
Department of Physics, University of Warwick, Gibbet Hill Road, Coventry
CV4 7AL, UK
 \and
Centre for Exoplanets and Habitability, University of Warwick, Gibbet Hill
Road, Coventry CV4 7AL, UK
 \and
Shanghai Observatory, Chinese Academy of Sciences, PR China
 \and 
 Purple Mountain Observatory, Chinese Academy of Science,
Nanjing 210008, PR China
 }

   \date{Received August 18, 2023; accepted March 12, 2024}

% \abstract{}{}{}{}{} 
% 5 {} token are mandatory
 
  \abstract
   {The \bpic{} system is the closest known stellar system with directly detected gas giant planets, an edge-on circumstellar disc, and evidence of falling sublimating bodies and transiting exocomets.
   The inner planet, \bpicc{}, has also been indirectly detected with radial velocity (RV) measurements.
   The star is a known \dsct{} pulsator, and the long-term stability of these pulsations opens up the possibility of indirectly detecting the gas giant planets through time delays of the pulsations due to a varying light travel time.
    We search for phase shifts in the \dsct{} pulsations consistent with the known planets \bpicb{} and c and carry out an analysis of the stellar pulsations of $\beta$ Pictoris over a multi-year timescale.
   We used photometric data collected by the BRITE-Constellation, bRing, ASTEP, and TESS to derive a list of the strongest and most significant \dsct{} pulsations.
   We carried out an analysis with the open-source python package \texttt{maelstrom} to study the stability of the pulsation modes of $\beta$ Pictoris in order to determine the long-term trends in the observed pulsations.
   We did not detect the expected signal for \bpicb{} or \bpicc{}.
   The expected time delay is 6 seconds for \bpicc{} and 24 seconds for \bpicb{}. 
    With simulations, we determined that the photometric noise in all the combined data sets cannot reach the sensitivity needed to detect the expected timing drifts.
    An analysis of the pulsational modes of $\beta$ Pictoris using \texttt{maelstrom} showed that the modes themselves drift on the timescale of a year, fundamentally limiting our ability to detect exoplanets around $\beta$ Pictoris via pulsation timing.}

   \keywords{Asteroseismology}

   \maketitle

\section{Introduction} \label{sec:Introduction}

\bpic{} is a nearby southern hemisphere star visible with the naked eye for which \dsct-like pulsations were discovered by \citet{Koen2003a}.
The planetary-mass companion \bpicb{} was detected using the VLT/NaCo instrument with direct imaging \citep{Lagrange2009b, Lagrange2010}.
Evidence of a second planet in the \bpic{} system was published by \citet{Lagrange2019b} using the radial velocity (RV) method and recently independently confirmed by \citet{Nowak2020} and \citet{Lagrange2020} using VLTI/GRAVITY observations.

The lifetime and frequency stability of \dsct{} pulsations make them astronomical ``stellar clocks,'' and therefore, they are great targets for applying timing techniques \citep{Compton2016}.
The common orbital motion of a star together with a companion around the barycenter of a system results in a periodic early or late arrival of the stellar pulsational signals.
This principle led to the first detection of planets orbiting a pulsar outside the Solar System \citep{Wolszczan1992, Wolszczan1994}.
The periodic variation of the arrival times can be seen as either a frequency modulation \citep[FM;][]{Shibahashi2012, Shibahashi2015} or a phase modulation \citep[PM;][]{Murphy2014, Murphy2015, Murphy2016b}.
The latter perspective works better for companions in wider orbits.

By applying the PM method on Kepler data, \citet{Murphy2016a} discovered a massive planet ($m \sin i \approx 12 M_J$) with an orbital period of about 840 days around a \dsct{} star.
In addition to the discovery of this planet, the PM method has led to the detection of 341 binaries and hundreds of more candidates \citep{Murphy2018a}.
Furthermore, it has provided us with the eccentricity, period, and mass function of these companions orbiting stars, just as the RV method does \citep[e.g. ][]{Nesvold2015}.
Applying the same method to pulsating stars observed by the TESS mission will lead to many more binary systems with full orbital solutions.

In this work, we use the data collected by the TESS satellite in its primary mission and data collected by the Hill sphere\footnote{The Hill sphere is the region around a planet where masses, such as moons and planetary rings, are gravitationally bound to the planet.} transit campaign, which was an international effort of space-based (e.g. through the BRITE-Constellation) and ground-based (e.g. through bRing, ASTEP) observations that searched for signatures of material around the giant planet \bpicb{} \citep{Kalas2019, Kenworthy2017}.
We analyzed this photometric data by searching for phase variations (and therefore time delays) caused by orbital motion in the pulsational signals.
\bpic{} was observed for approximately four months from October 2018 to February 2019 during the primary mission of TESS.
A second visit occurred during TESS' extended mission from November 2020 to February 2021 (see Table \ref{tab:obs} for a summary of all visits).

In Section~\ref{sec:The System}, we describe the properties of the different components in the \bpic{} system.
Section~\ref{sec:Observations} has a summary of all observational instruments and a frequency analysis for the photometry collected by TESS.
The theory and equations for this paper can be found in Section~\ref{sec:Methods}, and our results and conclusions follow in Sections~\ref{sec:Results} and \ref{sec:Conclusions}.

\section{The \bpic{} system}
\label{sec:The System}

\bpic{} (HD 39060; HR 2020) is one of the most studied and intriguing star-planet systems.
The Infrared Astronomical Satellite (IRAS) discovered an infrared excess \citep{Aumann1984} for this bright and close southern star that was attributed to the presence of a circumstellar disc.
This disc was first imaged by \citet{Smith1984} and clearly showed the edge-on geometry of this system.
The gas and dust in this disc is mostly ``second generation,'' that is, constantly replenished by collisions of comets and asteroids \citep{Lagrange2000}.
A warp in this disc \citep{Augereau2001, Mouillet1997, Nesvold2015} and signatures of evaporating exocomets (also called falling evaporating bodies, or FEBs) in spectroscopy \citep{Ferlet1987, Beust2000} were attributed to an exoplanet orbiting the star interacting dynamically with its environment  \citep[for more information on exocomets observed around \bpic{} detected in photometry, see][]{Zieba2019, Strom2020, Pavlenko2022, LecavelierdesEtangs2022}.

\subsection{The star}
\label{sec:The star}

\citet{Koen2003a} discovered \dsct{}-type pulsations at the millimagnitude level originating from \bpic.
Further analysis by \citet{Mekarnia2017}, \citet{Zwintz2019}, and \citet{Zieba2019} showed dozens of additional frequencies between 20 and 80 cycles per day.
An asteroseismic large spacing, $\Delta\nu$, has been measured for \bpic{} \citep{Bedding2020}, which might facilitate a precise asteroseismic age in the future.
The pulsations also induce intrinsic variations in the RV at $\lesssim$ 1 \kms{} peak to peak \citep{Lagrange2009a,Lagrange2012,Galland2006}, which hampers the search for planets with the RV method in this system. 
A selection of the fundamental properties of the star \bpic{} are listed in Table \ref{tab:bpic}.

\begin{table}[ht]
\centering
 \caption{Various stellar parameters of the star \bpic.}\label{tab:bpic}
\renewcommand{\arraystretch}{1.35}
\begin{tabular}{lcc}
 \hline \hline
 Parameter &
 Value &
  Reference
\\ \hline
RA (J2000.0)        & 05h 47m 17.09s            & 1 \\ 
DEC (J2000.0)       & -51h 03m 59.41s           & 1 \\
V (mag)             & 3.86                      & 2 \\
TESS (mag)          & 3.696                     & 1 \\
age (Myr)           & 23 $\pm$ 3            & 3 \\
parallax (mas)      & 50.93 $\pm$ 0.15                 & 4,5,6 \\
Distance (pc)       &  19.63 $\pm$ 0.06                & 4,5,6 \\
Spectral class      & A6V                       & 7 \\
Radius (\Rsun)      & 1.497 $\pm$ 0.025                 & 8 \\
Mass (\Msun)        & 1.75$_{-0.02}^{+0.03}$    & 9 \\
Teff (K)            & 8090  $\pm$ 59         & 8 \\
\hline
\end{tabular}
\tablebib{(1)~\citet{Stassun2019};
(2) \citet{Cousins1971}; 
(3) \citet{Mamajek2014};
(4) \citet{Gaia2016};
(5) \citet{Gaia2023};
(6) \citet{Lindegren2021};
(7) \citet{Gray2006};
(8) \citet{Zwintz2019};
(9) \citet{Lacour2021}.
}
\end{table}

\subsection{The planets: \bpicb{} and c}
\label{sec:The planet}

The warp of the inner disc of \bpic{} observed by the Hubble Space Telescope and in ground-based observations was one of the indirect lines of evidence for a massive substellar companion orbiting the star \citep{Burrows1995, Mouillet1997, Heap2000, Golimowski2006}.
Signatures of infalling exocomets in the spectra of the star also needed a ``perturber'' to scatter them onto eccentric inner system-bearing orbits.
The planet, \bpicb, was then discovered using the VLT/NaCo instrument data in 2003 \citep{Lagrange2009b} and was later confirmed by \citet{Lagrange2010}.
A transit-like event was observed in 1981 and attributed to a planet \citep{Lecavelier1995}. However, a better orbit determination with the VLT/SPHERE instrument ruled out \bpicb{} as the cause of that event \citep{Lagrange2019a}.
Using data from the Gemini Planet Imager, \citet{Wang2016} were able to rule out a transit of the planet \bpicb{} during the conjunction in 2017 at a 10$\sigma$ level.
However, a Hill sphere transit was predicted for the time between late 2017 and early 2018 \citep{Lecavelier2016, Wang2016}.
Various observational campaigns were initiated by PicSat \citep{2018SPIE10698E..21N,2022PASP..134c4501M}; bRing \citep{Kenworthy2017}; and the BRITE-Constellation \citep{Weiss2014} in order to photometrically observe possible material around the planet; however, there was no significant detection \citep{Kenworthy2021}.
The mass and the orbital solution of the outer planet \bpicb{} are listed in Table \ref{tab:bpicb}.

\begin{table}[ht]
\centering
 \caption[]{Various parameters of the planets \bpicb{} and c based on \citet{Lacour2021}.}\label{tab:bpicb}
\renewcommand{\arraystretch}{1.35}
\begin{tabular}{lccc}
 \hline \hline
 Parameter & Unit &
 \bpicb &
  \bpicc
\\ \hline
Mass     & \MJ & $11.90^{+2.93}_{-3.04}$      & $8.89^{+0.75}_{-0.75}$ \\ 
a        & au & $9.93^{+0.03}_{-0.03 }$      & $2.68^{+0.02}_{-0.02}$ \\
e        & --- & $0.103^{+0.003}_{-0.003}$    & $0.32^{+0.02}_{-0.02}$ \\
i        & $^{\circ}$  &$89.00^{+0.00}_{-0.01}$      & $88.95^{+0.09}_{-0.10}$ \\
$\varpi$\tablefootmark{(a)} & $^{\circ}$    & $199.3^{+2.8}_{-3.1}$ & $66.0^{+1.8}_{-1.7}$ \\
$\tau$\tablefootmark{(b)}   & ---  & $0.719^{+0.008}_{-0.010}$    & $0.724^{+0.006}_{-0.006}$ \\
P        & years         & $23.61^{+0.09}_{-0.09}$ & 3.34  $\pm$ 0.04 \\
P        & days         & $8623^{+31}_{-32}$       & 1221 $\pm$ 15 \\
$t_{p}$\tablefootmark{(c)}    & MJD        & 65243                   & 59888 \\
\hline
\end{tabular}
\tablefoot{\\
\tablefoottext{a}{The argument of periastron $\varpi$ reported in \citet{Lacour2021} follows the definition in \citet{Blunt2020} and therefore refers to the orbit of the companion and not to the star. In this paper, however, we refer to the orbit of the star when we use $\varpi$.}\\ 
\tablefoottext{b}{reference epoch MJD 59000 (31 May 2020).}\\
\tablefoottext{c}{derived from $\tau$.}
}

\end{table}

Evidence of an additional planet in the \bpic{} system was published by \citet{Lagrange2019b}.
Over 6000 spectra of the star taken between 2003 and 2018 by the HARPS instrument at the ESO La Silla 3.6 m telescope have been analyzed, and they showed a hint of a planetary signal.
\bpicc{} was then ultimately directly detected by \citet{Nowak2020} and \citet{Lagrange2020} using VLTI/GRAVITY observations.
A list of parameters for the planet can be found in Table \ref{tab:bpicb}.

\section{Observations}
\label{sec:Observations}

Due to the 2017-2018 Hill Sphere Transit of \bpicb, an international campaign of space- and ground-based observations was launched in order to search for signatures of material around the giant planet \citep{Kalas2019,Kenworthy2021}.
Table \ref{tab:obs} summarizes various properties of the different light curves.
Changes to those light curves other than the Gaussian high-pass procedure, which is explained in Section \ref{sec:Gaussian Highpass filter}, are noted in the following subsections.
The data provided by BRITE-Constellation was left unchanged.
A detailed analysis of the photometry of \bpic{} collected by BRITE-Constellation and bRing was published by \citet{Zwintz2019}.
As we wanted to measure the periodic motion of a star around a barycenter, we also wanted to correct for the motion the Earth in the Solar System.
All the observations in this work were therefore converted to the Barycentric Julian Date in the barycentric dynamical time (BJD$_\text{TBD}$) standard using the {\tt Python} tool {\tt BARYCORRPY} \citep{Kanodia2018}, which is based on the IDL code {\tt BARYCORR} \citep{Wright2014}.

\begin{table*}[ht]
\centering
 \caption[]{Summary of the properties of the various instruments and corresponding light curves.}\label{tab:obs}
\renewcommand{\arraystretch}{1.35}
\begin{tabular}{lllllllll}
\hline\hline
Observation & \begin{tabular}[c]{@{}l@{}}Wavelength \\ (nm)\end{tabular} & \begin{tabular}[c]{@{}l@{}}Observation  \\ start \end{tabular} & \begin{tabular}[c]{@{}l@{}}Observation  \\ end \end{tabular} & \begin{tabular}[c]{@{}l@{}}$T$ \\ (days)\end{tabular} & \begin{tabular}[c]{@{}l@{}}$1/T$\\ ($10^{-3}$ \cd) \end{tabular}  & \begin{tabular}[c]{@{}l@{}}$f_\text{Ny.}$  \\ (\cd) \end{tabular}     & \begin{tabular}[c]{@{}l@{}}cadence  \\ (s) \end{tabular} &\begin{tabular}[c]{@{}l@{}}DC  \\ (\%) \end{tabular} \\\hline
BHr         & 550 - 700                                                  & 16 March 2015 & 2 June 2015   & 78.32    & 12.77                          & 4167  & 10.37 & 6.78   \\
BTr + BHr   & 550 - 700                                                  & 4 Nov. 2016    & 17 June 2017  & 224.6    & 4.453                          & 2128  & 20.30 & 7.07   \\
BHr         & 550 - 700                                                  & 9 Nov. 2017    & 25 April 2018 & 167.3    & 5.976                          & 2128  & 20.30 & 7.48   \\\hline
bRing       & 463 - 639                                                  & 2 Feb. 2017    & 1 Sept. 2018   & 575.5    & 1.738                          & 135.4 & 319.1 & 27.0   \\\hline
ASTEP17     & 695 - 844                                                  & 28 March 2017 & 14 Sept. 2017  & 170.0    & 5.881                          & 495.8 & 87.13 & 18.9   \\
ASTEP18     & 695 - 844                                                  & 28 March 2018 & 15 July 2018  & 109.3    & 9.150                          & 502.8 & 85.92 & 29.2   \\\hline
TESS        & 600 - 1000                                                 & 19 Oct. 2018   & 1 Feb. 2019    & 105.2    & 9.507                          & 360.0 & 120.0 & 85.3   \\
TESS        & 600 - 1000                                                 & 20 Nov. 2020   & 8 Feb. 2021    & 79.8    & 12.53                          & 360.0 & 120.0 & 90.2   \\\hline
\end{tabular}
\tablefoot{\\
The term $T$ denotes the time base of the observations, the reciprocal value $1/T$ corresponds to the Rayleigh criterion, $f_\text{Ny.}$ is the Nyquist frequency, and DC is the duty cycle.
BRITE Lem (BLb) is equipped with a blue filter and observed \bpic{} from December 2016 until June 2017, but due to significantly higher noise in the time series, the data was disregarded from the analysis. See \citet{Zwintz2019} for an analysis of the BLb observations.
}

\end{table*}

\begin{figure}[ht!]
\centering
\includegraphics[trim=0.3cm 0cm 1.7cm 0cm, clip,width=1\columnwidth]{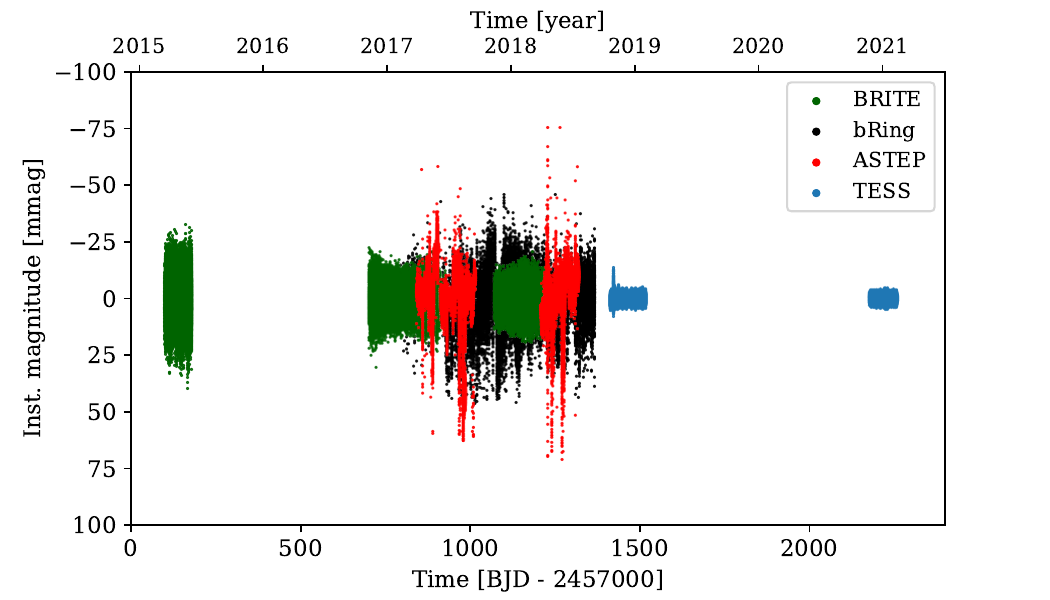}
\caption{Full light curve of all available observations of the star \bpic{} used in this work.}
\label{fig:total_lc}
\end{figure}

\subsection{BRITE-Constellation}

The BRITE-Constellation \citep{Weiss2014} consists of five nanosatellites collecting photometry for the brightest stars on the sky.
In this work, we analyzed data collected by three of the satellites: BRITE-Heweliusz (BHr), BRITE-Toronto (BTr), and BRITE-Lem (BLb).
Being in a low-earth orbit, the orbital periods of the satellites are all around 100 minutes.
A minimum of 15 minutes per orbit is dedicated to observations.
Three different runs were conducted in the constellations around Pictor and Vela, which also included the star \bpic.
A summary of the durations and various properties of the observations can be found in Table \ref{tab:obs}.
The pipeline for the photometry reduction is described in \citet{Popowicz2017}.
An analysis of all BRITE observations was conducted in \citet{Zwintz2019}.
For the three runs by BHr, BTr+BHr, and BHr, which all used the red BRITE filter, six, 13, and eight significant frequencies were extracted, respectively.
The only run with a blue filter by BLb suffered from higher noise compared to the other BRITE observations.
\citet{Zwintz2019} has reported four frequencies in the collected BLb photometry.
The blue observations were discarded from this analysis, as the data quality was not good enough to provide additional information.

\subsection{bRing}\label{subs:bring}

The bRing project (which stands for ``the \bpicb{} Ring project'') was initiated in order to collect photometry of \bpic{} during the Hill sphere transit of \bpicb{} at the end of 2017 \citep{Stuik2017}.
To that end, two stations in South Africa and Australia were built, each consisting of two wide-field cameras.
Their design is based on the Multi-Site All-Sky CAmeRA (MASCARA) \citep{Snellen2012, Talens2017}.
The capability of bRing to monitor bright stars and find previously unknown variables has been shown by \citet{Mellon2019}.
More information on the observing strategy and design of bRing can be found in \citet{Stuik2017}.
The reduction pipeline for the MASCARA and bRing instruments is described in \citet{Talens2018}. 
With a passband of 463 -- 639 nm, bRing collected the shortest wavelengths of all observatories considered in this work. We expected to see the highest pulsational amplitudes in these data, as \bpic{} is a star of spectral type A6 \citep{Zwintz2019} and has its energy maximum in the blue optical wavelengths.
Due to some evident outliers in the data, one 5$\sigma$ clip with respect to the median of the data set was applied.
This significantly weakened the one-day aliases in the spectral window.
An iterative sigma clipping procedure was not conducted due to noticeable changes in the amplitudes of the pulsations in this case (see \citet{Hogg2010} for a discussion of sigma clipping in order to remove outliers).
The observations by bRing were separated into two equally sized segments in order to gain more time delay measurements while also maintaining a precision in frequency and phase comparable to the ASTEP observations. 
\citet{Zwintz2019} found six significant frequencies in the photometry collected by bRing.
All of them are also identified in the data collected by BRITE, ASTEP, and TESS.

\subsection{ASTEP}

The Antarctic Search for Transiting ExoPlanets, or ASTEP, is an automated telescope with an aperture of 40 cm located at the Concordia station at Dome C in Antarctica \citep{Abe2013, Guillot2015, Mekarnia2017}. It uses a Sloan i' filter (centered at 763 nm). 
We only used measurements with a sun elevation lower than -18$^{\circ}$.
Notably, data points where the centroid of the star did not fall on the central pixel suffer from strong outliers.
The removal of these outliers and a 5$\sigma$ clip with respect to the median weakened aliases significantly but without noticeable changes in the amplitude of the strongest pulsational frequencies.
\citet{Mekarnia2017} conducted a frequency analysis of the \bpic{} photometry collected by the ASTEP observatory, and they are consistent with the ones seen in the TESS data.

\subsection{TESS}

The Transiting Exoplanet Survey Satellite \citep[TESS;][]{Ricker2015} was launched in April 2018 in order to find transiting exoplanets around nearby bright stars.
The data of \bpic{} (TIC 270577175, T = 3.696 mag) was collected from 19 October 2018 to 1 February 2019 in sectors four through seven and from 20 November 2020 to 8 February 2021 in the sectors 32 through 34.
The data for first four sectors were obtained during TESS' primary mission, and data for the three other sectors were obtained approximately two years later as part of the first extended mission.
\bpic{} is one of the preselected targets for which short-cadence (2 minutes) data are provided.
Due to the high-cadence data, the high photometric precision of TESS, its high duty cycle, and the long baseline, \dsct{} pulsations can be resolved and identified with high precision.
The photometric data of \bpic{} as observed by TESS was accessed and modified with the Python package {\tt lightkurve} \citep{Lightkurve2018}, which retrieves the data from the MAST archive.\footnote{\url{https://archive.stsci.edu/tess/}}
For this analysis, we used the Pre-search Data Conditioning Simple Aperture Photometry \citep[PDCSAP;][]{Smith2012, Stumpe2012} light curves, which are produced from the Science Processing Operations Center (SPOC) pipeline \citep{Jenkins2016, Jenkins2017, Jenkins2010}.
These PDCSAP light curves were corrected for systematics by the SPOC pipeline. 
We also visually inspected the target pixel files (TPF) in order to rule out various instrumental and astrophysical effects, such as Solar System asteroids or comets crossing the field of view. 
A comparison of the Lomb-Scargle periodogram \citep{Lomb1976, Scargle1982} of the raw simple aperture photometry and PDCSAP light curves showed a significant change in the noise at low frequencies.
This is due to the systematic effects present in sector four. 
The lowest noise in the low-frequency range can be found for the PDCSAP light curve with a completely removed fourth sector.
This light curve was then used for the main frequency analysis.

The individual sectors were normalized by dividing each of the sectors by their respective median flux, and the sectors were combined into one light curve.
Furthermore, every measurement with a non-zero quality flag (see Sect. 9 in the TESS Science Data Products Description Document)\footnote{\url{https://archive.stsci.edu/missions/tess/doc/EXP-TESS-ARC-ICD-TM-0014.pdf}} was removed.
Such anomalies as cosmic ray events or instrumental issues were marked by these quality flags.
 
The frequency analysis was conducted using the Python package {\tt SMURFS} \citep{Mullner2020} and checked with the software package {\tt Period04} \citep{Lenz2005}.
Following \citet{Breger1993}, all pulsation frequencies down to a signal-to-noise ratio of four were extracted.
The frequency range analyzed is between zero and the Nyquist frequency of 360 cycles per day.
Following the procedure described in \cite{Zieba2019}, 37 significant p-modes in frequencies ranging from 34 to 76 $\text{d}^{-1}$ were identified.
As we are only interested in the strongest pulsational frequencies for this time delay analysis, we did not try to further recover any of the lower amplitude modes. 
A list of the extracted frequencies can be found in Table \ref{tab:freqs}.

\section{Theory and methodology}
\label{sec:Methods}

In this chapter, we discuss the theory behind time delays and the methods used in order to finally arrive at the time delay plot. Importantly, this plot can be used to search for companions around pulsating stars. 

\subsection{\dsct{} stars}

\dsct{} stars can be found at the intersection region between the main sequence and the instability strip on the Hertzsprung-Russel diagram.
Thanks to the nearly uninterrupted, high-precision photometry of Kepler's primary four-year mission, the general understanding of pulsating stars has been revolutionized.
\dsct{} stars have masses between approximately 1.5 and 2.5 \Msun.
They pulsate in radial and non-radial low-degree, low-order pressure (p) modes that are primarily driven by an opacity mechanism (also called a $\kappa$-mechanism) in their HeII zone with contributions from turbulent pressure \citep{Houdek2000,Antoci2014} and the edge-bump mechanism \citep{Murphy2020}.
The oscillations have periods between 18 minutes and 8 hours respectively 80 and three cycles per day \citep{Aerts2010}.
Linear combinations of those oscillations can, however, create peaks at lower frequencies \citep{Breger2014}.
Besides main-sequence and more-evolved stars, \dsct{} pulsations were observed in pre-main-sequence stars, thus giving us the possibility to learn about early stellar evolution \citep{Zwintz2014,Murphy2021,Steindl2022}.

\subsection{The ephemeris equation}

The search for time delays in certain astrophysical signals requires a (quasi-)periodic process in space.
A review on this and the related equations can be found in \citet{Hermes2018}.
There are different processes that are ``clock-like'' under the assumption of a closed system, including the exceptionally stable signals of pulsars, the eclipse time of binary stars, or certain pulsating stars, as in our case.
Deviations from periodic signals can be used to analyze the spin-down of pulsars or to discover companions around pulsars \citep{Wolszczan1992, Wolszczan1994}, eclipsing binaries \citep{Barnes1975}, or pulsating stars \citep{Silvotti2007} \citep[for a general review of pulsating stars in binary systems see][]{Murphy2018b}.
To do so, one creates O-C (observed minus calculated) diagrams \citep[see e.g.][]{Sterken2005} in order to search for deviations from the predicted ephemeris in the observations.
O-C diagrams work the best if the star is pulsating in only a single mode and if the maxima are narrow and well defined, as they are easy to track in that case.
However, these diagrams especially struggle with multi-mode pulsators. 

\subsection{Frequency modulation and phase modulation: The state of the art}
\label{sec:FM and PM}

Building on the established methods of O-C diagrams, two new and complementary techniques have emerged for finding companions around pulsating stars.
The FM method \citep{Shibahashi2012} searches and analyzes the variations in the frequency of a pulsating star induced by a companion.
The periodic FM creates multiples around every pulsation peak in the frequency spectrum.
Their frequencies, relative amplitudes, and phases can be used to get a full orbital solution, as described in \citet{Shibahashi2015}.
The effectiveness of the FM method was validated through a comparison with an eclipsing binary system \citep{Kurtz2015},
and it is best suited for data sets with a baseline that exceeds the orbital period of the companion.

The PM method is more sensitive to companions in wider orbits. It was developed by \citet{Murphy2014}, \citet{Murphy2015}, and \citet{Murphy2016b}.
\citet{Compton2016} showed that \dsct{} stars and white dwarfs are best suited for this method.
Its effectiveness was demonstrated by \citet{Schmid2015} by showing the binary nature of KIC10080943 using the PM method and attributing certain pulsations to the corresponding star in the binary due to the antiphase modulation in time delays.
Such a system with observable time delays in both components is called a PB2, analogous to spectroscopic terminology, where binary star systems are called SB2s if both stars show observable RVs.
Other proof of the functionality of the PM method was shown by \citet{Derekas2019} by comparing the orbital parameters derived from RV with those from PM.

An additional advantage of the PM method is its easier automation for many stars.
When applying this method to 2224 main-sequence A/F stars in the four-year main Kepler data, \citet{Murphy2018a} were able to find 317 PB1 systems, where only one component is pulsating and showing time delays, and 24 PB2 systems, where two stars are pulsating.
It is worth noting here that determining orbital solutions using spectra and generating RV curves for the same number of stars would be much more time intensive.

Other methods were developed by \citet{Koen2014} and \citet{Balona2014} to search for binary systems by tracing the \dsct{} pulsations of stars.
In contrast to the FM and PM methods, these methods are not able to provide a full orbital solution, which is usually gained by analyzing RV curves of spectroscopic binaries.  

\subsection{Time delays}
\label{sec:Time Delays}

Time delays arise when a signal (in our case always an electromagnetic wave with the propagation velocity defined by the speed of light) has to travel different distances at different epochs.
Following \citet{Smart1977} and \citet{Balona2014}, the distance $r$ between the pulsating star and the center of gravity of its system can be described by

\begin{equation} 
\label{eq:r}
r = \frac { a _ { 1 } \left( 1 - e ^ { 2 } \right) } { 1 + e \cos f },
\end{equation}

\noindent
where $a_1$ denotes the semi-major axis of the star, $e$ is its eccentricity, and $f$ is the true anomaly.
The distance to the star varies relative to the Earth by

\begin{equation} 
\label{eq:z}
z =  r \sin ( f + \varpi ) \sin i,
\end{equation}

\noindent
with $\varpi$ being the argument of periapsis, that is, the angle between the nodal point and the periapsis,\footnote{The argument of periapsis is usually denoted with $\omega$.
This symbol, however, is used in asteroseismology to denote the angular oscillation frequency.
Also, one should not confuse $\varpi$ with the longitude of periapsis, which is the sum of the longitude of the ascending node $\Omega$ and the argument of periapsis.} and $i$ as the inclination of the system.

At this point, Equation \ref{eq:r} can be substituted into Equation \ref{eq:z}.
The time delay \(\tau = - z/c \) is then completely described by the following equation:

\begin{equation} 
\label{eq:tau}
 \tau(t,\textbf{x}) = - \frac{a_1 \sin i}{c} (1-e^2) \frac{\sin f \cos \varpi + \cos f \sin \varpi}{1 + e \cos f}.
\end{equation}

The set \( \textbf{x} = (\Omega = 2 \pi / P, a_1 \sin i / c, e, \varpi, t_p) \) in Equation \ref{eq:tau} includes all of the system-specific parameters needed to describe the time delay for a given time $t$.
The term $P$ is the orbital period of the system, or equivalently $1/P = \nu_\text{orb}$ the orbital frequency, and thus $\Omega$ is the angular orbital frequency.
The projected semi-major axis of the pulsating star is described by $a_1 \sin i$.
Dividing this quantity by the speed of light $c$ gives us the size of the orbit for the pulsating star in light seconds.
The argument of periapsis is described by $\varpi$ and the time of periapsis passage by $t_p$.
(For a graphical visualization of the orbital parameters, see \citet{Murphy2015}.)

The two trigonometric functions of the true anomaly, $\sin f$ and $\cos f$, can be expressed in terms of series expansions and Bessel functions:

\begin{equation} 
\label{eq:cosf}
\cos f = - e + \frac { 2 \left( 1 - e ^ { 2 } \right) } { e } \sum _{ n = 1 } ^ { \infty } J_{ n } ( n e ) \cos n \Omega \left( t - t _ { \mathrm { p } } \right),
\end{equation}

\begin{equation} 
\label{eq:sinf}
\sin f = 2 \sqrt { 1 - e ^ { 2 } } \sum_{ n = 1 } ^ { \infty } J_{ n } ^ { \prime } ( n e ) \sin n \Omega \left( t - t _ { \mathrm { p } } \right),
\end{equation}

\noindent
with \(J_n'(x) = \text{d}J_n(x) / \text{d}x\) \citep[the derivation of Equation \ref{eq:cosf} and \ref{eq:sinf} can be found in Appendix A of][]{Shibahashi2015}.
The changing distances between us and the clock in space are fundamentally connected with varying radial velocities, $v_{\text {rad } }$:

\begin{equation} 
\label{eq:vrad}
 v_\text{rad} = c \frac{\text{d}\tau}{\text{d}t}.
\end{equation}

By substituting Equation~\ref{eq:tau} into Equation~\ref{eq:vrad}, we obtain

\begin{equation} 
\label{eq:vrad_all}
v_{ \text { rad } } = - \frac { \Omega a_{ 1 } \sin i } { \sqrt { 1 - e ^ { 2 } } } [ \cos ( f + \varpi ) + e \cos \varpi ].
\end{equation}

Given Equation~\ref{eq:vrad} and the convention that a positive RV corresponds with a receding object and a negative RV with an approaching one, we could deduce the following: a negative time delay is due to an early arrival of the signal, that is, the star is closer to us, and vice versa\footnote{This convention for the time delays was established with \citet{Murphy2015}, while \citet{Murphy2014} used reversed signs;their plots are therefore mirrored around the vertical axis.} (see Table \ref{tab:sign}). 

\begin{table}[ht!]
\centering
\caption{Sign convention for the RV $v_{\text{rad}}$ and the time delays $\tau$.}
\label{tab:sign}
\begin{tabular}{l|cc}
\multicolumn{1}{c|}{} & positive sign (+)   & negative sign (-)                           \\ \hline
$v_\text{rad}$             & moving away            & approaching                 \\
$\tau$                & farther away / late arrival& closer / early arrival
\end{tabular}
\end{table}

One can see in Equations~\ref{eq:tau} and \ref{eq:vrad_all} that the time delay as well as the RV of a system can be completely described by the orbital parameters.
If we obtain those parameters by one method, we can predict what we should observe with the other one.
Furthermore, if we generate the time delay plot from our observations, we can apply a chi-squared minimization technique in order to get the parameters in set $\textbf{x}$.
This concept was introduced with \citet{Murphy2015} and is a major improvement to \citet{Murphy2014}, where the time delay measurements were numerically differentiated in order to derive the parameters from the obtained RV curve.

Finally, by using two of the derived orbital parameters, \(a_1 \sin i / c \) and $P_\text{orb}$, we can calculate the mass function \(f \left( m_{ 1 } , m_{ 2 } , \sin i \right)\) for the binary system:

\begin{equation} 
f \left( m_{ 1 } , m_{ 2 } , \sin i \right) : = \frac { \left( m_{ 2 } \sin i \right) ^ { 3 } } { \left( m_{ 1 } + m_{ 2 } \right) ^ { 2 } } = \frac { 4 \pi ^ { 2 } c ^ { 3 } } { G } v_{ \text { orb } } ^ { 2 } \left( \frac { a_{ 1 } \sin i } { c } \right) ^ { 3 },
\end{equation}

\noindent
where $m_2$ is the mass of the (usually non-pulsating) companion and $G$ is the gravitational constant.

\subsection{Phase modulation method: Methodology}
\label{sec:The Phase Modulation Method}

Before we could create the time delay plot, we had to analyze the change in the phase of the various pulsation modes with time.
The basic equations for that can be found in \citet{Murphy2014} and are summarized in the following.
We started by dividing the light curve into $n$ equally sized segments.
Then, we calculated the phase in every segment for each frequency.
This left us with a series of phases $\Phi_{ j }$ for every segment $(1, 2,\ldots, n)$ for a fixed frequency $\nu_j$:
 
\begin{equation} 
\Phi_{ j } = \left[ \phi_{ 1 j } , \phi_{ 2 j } , \ldots , \phi_{ i j } , \ldots , \phi_{ n j } \right].
\end{equation}

Numerically, the phase in a segment is derived by calculating the argument of the Fourier Transformation in the respective segment:

\begin{equation} 
\Phi ( t ; \nu ) = \tan ^ { - 1 } \left( \frac { \operatorname { Im } ( \mathrm { F } ( t ; \nu , \delta t ) ) } { \operatorname { Real } ( \mathrm { F } ( t ; \nu , \delta t ) ) } \right),
\end{equation}

\noindent
where \( \mathrm { F } ( t ; \nu , \delta t ) \) is the value of the Fourier Transformation of the time series for frequency $\nu$ in segment $\delta t$.

As phases are frequency dependent, the resulting phase shifts have different amplitudes for different frequencies. 
To get rid of this effect, we converted them into time delays by first calculating the relative phase shifts:

\begin{equation} 
\Delta \phi_{ i j } = \phi_{ i j } - \overline { \phi }_{ j },
\end{equation}

\noindent
with $ \overline { \phi }_{ j }$ as the mean phase of frequency $\nu_j$:

\begin{equation} 
 \overline { \phi }_{ j } = \frac { 1 } { n } \sum_{ i = 1 } ^ { n } \phi _ { i j }.
\end{equation}

The time delay $\tau_{ij}$ for segment $i$ and frequency $\nu_j$ is thus simply the relative phase shift divided by the angular pulsation frequency:

\begin{equation} 
\label{eq:TD_ij}
\tau_{ij} = \frac { \Delta \phi_{ij} } { 2 \pi \nu_j }.
\end{equation}

A planet with an orbital frequency of $v _ { \mathrm { orb }}$ in a circular orbit will induce time delays that can be described by a sine function with phase $\psi$:

\begin{equation} 
\tau ( t ) = A \sin \left( 2 \pi v _ { \mathrm { orb } } t + \psi \right).
\end{equation}

\begin{figure}[ht!]
\centering
\includegraphics[width=0.5\textwidth]{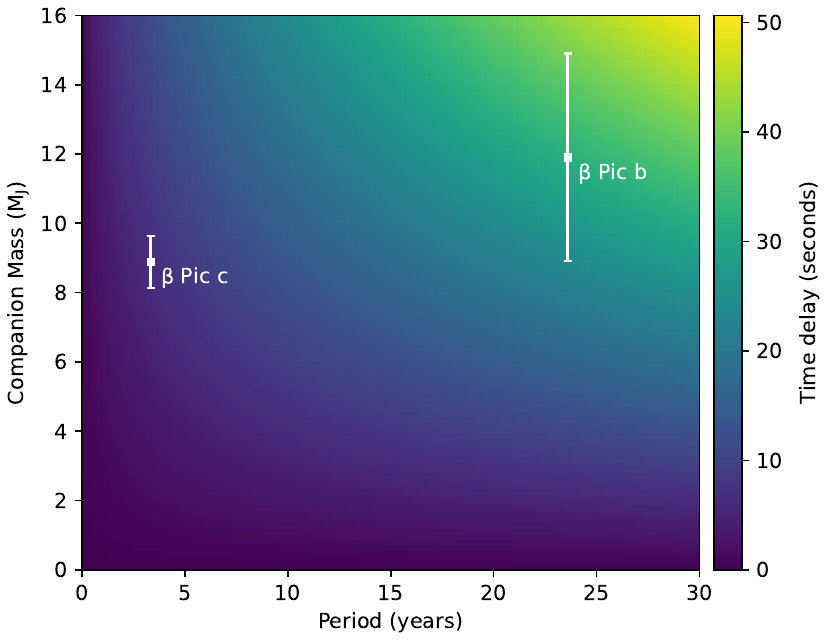}
\caption{Time delays for the \bpic{} system. 
The colors indicate the expected time delays for an edge-on planet in a circular orbit.
The uncertainties in the orbital period for the planets are smaller than the marker size.}
\label{fig:TD_period}
\end{figure}

The amplitude $A$ can be simply derived by inserting the center of mass equation \( m_1 a_1 = m_2 a_2\) into \( \tau = a_1 \sin i / c\):

\begin{equation} 
\label{eq:TD_period}
\tau  = \frac{a_\text{P} \sin i}{c} \frac{M_\text{S}}{M_*},
\end{equation}

\noindent
with $a_\text{S}$ as the semi-major axis of the companion. The terms $M_\text{S}$ and $M_*$ are the mass of the companion and the star, respectively.
Using Equation~\ref{eq:TD_period} and the mass of around 1.8 \Msun{} given in Table \ref{tab:bpic}, we could calculate the expected time delay for a given period.
This is visualized in Figure \ref{fig:TD_period}.
The time delay is around 24 seconds for \bpicb{} and 6 seconds for \bpicc{}.
For comparison, the smallest time delay detected in the main Kepler data is 7 seconds \citep{Murphy2016a}.

For more eccentric orbits, the pulsation time plot is described by a sum of harmonics with amplitudes $A_k$ and phases $\phi_k$ corresponding to order $k$:

\begin{equation} 
\tau ( t ) = \sum_{ k = 1 } ^ { N } A_{ k } \sin \left( 2 \pi k v_{ \mathrm { orb } } t + \psi_{ k } \right).
\end{equation}

\begin{figure}[ht!]
\centering
\includegraphics[width=0.5\textwidth]{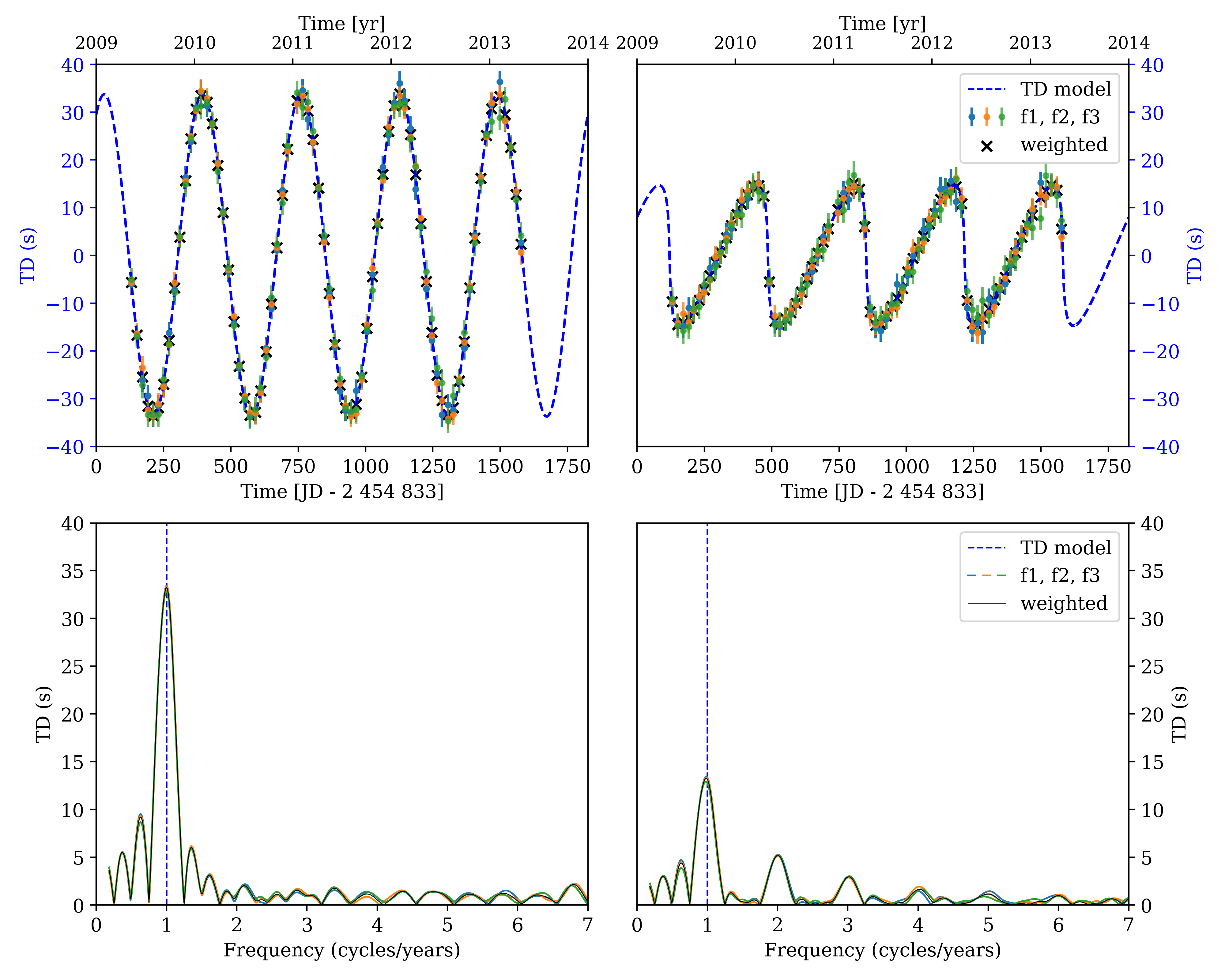}
\caption{Simulation of a companion in a circular (\(e = 0\); left column) and eccentric (\(e = 0.9\); right column) orbit as if observed by the Kepler Space Telescope.
The following parameters were used: \(P = 1\) year, $\varpi = 0$, \(M_\text{pulsating} = 1.8 \) \Msun{}, and \(M_\text{companion} = 0.1 \) \Msun.
This led to a semi-amplitude of around 34 seconds in the circular case (using Equation \ref{eq:TD_period}) and around 15 seconds in the eccentric case (using Equation \ref{eq:TD_eccentric}). \textit{Upper panel:} Simulated time delay plot.
\textit{Lower panel:} Fourier transformation of the time delays.
One can clearly see the relative increase of the first harmonic at two cycles/year for the eccentric case.}
\label{fig:TDsim}
\end{figure}

The height of the first harmonic relative to the one of the orbital frequency is a measure of the eccentricity.
The theory behind this is described in Appendix A of \citet{Murphy2014}. 
A visualization of that can be seen in Figure \ref{fig:TDsim}.
An increase in eccentricity also influences the amplitude of the time delay. This is given by the following equation:

\begin{equation} 
\label{eq:TD_eccentric}
\frac{a_1 \sin i}{c} = \frac{(\tau_{\text{max}} - \tau_{\text{min}})}{2} \left( 1 - e^2 \cos^2 \varpi \right)^{-1/2}.
\end{equation}

The maximum time delay is therefore reached in the case of $\varpi = \pm \pi / 2$ or for the simple circular orbit case.

The larger the ratio between the orbital size $a_1 \sin i / c$ and the pulsation period $1/\nu$, the higher the sensitivity of the method \citep{Murphy2016b}.

Due to the size of the segments, one has to make a trade-off between time or frequency resolution. Using a shorter segment size has the advantage of a finer sensitivity at periastron; however, the uncertainties are simultaneously increased because of a poorer frequency resolution in the Fourier transform.

Under the assumption of Gaussian noise, increasing the cadence of an observation by a factor of $N$ decreases the uncertainties in the measured phases by a factor of $\sqrt{N}$ \citep{Murphy2012}. 
The phase errors also scale inversely with amplitude, which means that the most valuable frequencies are the ones with the highest amplitudes.

\subsection{Intrinsic amplitude and phase variations}
\label{sec:4.7}

Amplitude modulations in \dsct{} stars have been observed in the past and thoroughly analyzed in the four-year main Kepler data by \citet{Bowman2016}. 
Additionally, due to intrinsic reasons for those modulations (such as the coupling of pulsational modes or pairs of close unresolved frequencies leading to a beating effect), binarity can cause variability. 

\bpic{} is known to show amplitude variation in certain pulsational frequencies, as reported by \citet{Zwintz2019} and \citet{Mekarnia2017}. However, PMs have not been observed yet \citep{Zwintz2019}.

\subsection{Light curve reduction}

\begin{figure}
\centering
\includegraphics[width=0.5\textwidth]{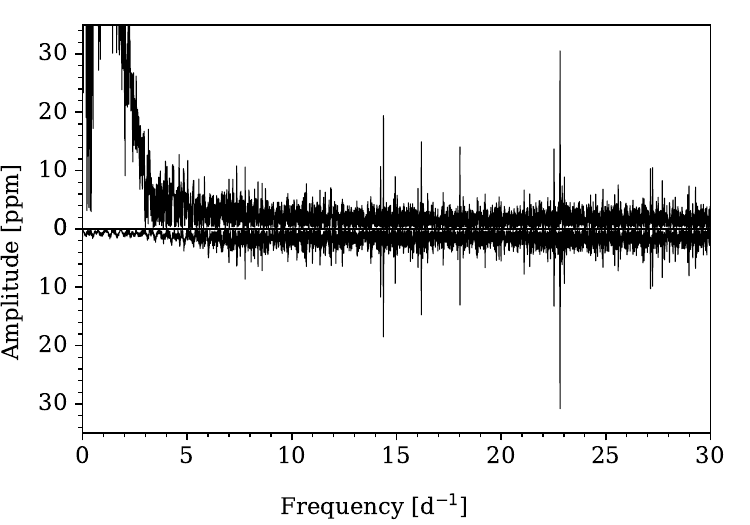}
\caption{Comparison of the amplitude spectra of the ``raw'' PDCSAP light curve (upper panel) and the Gaussian high-pass filtered light curve (lower panel).
The power of the peaks below five \cd{} are significantly weakened without influencing the \dsct{} pulsations.}
\label{fig:RawvsGauss1}
\end{figure}

Following \citet{Murphy2016a}, unused frequencies were pre-whitened from our light curves, as their presence adds unwanted variance to the data. 
Furthermore, a high-pass filter was applied to the light curve to remove any remaining instrumental signal and low-frequency oscillations, preserving all content at frequencies above 5 \cd. 
The effect of a high-pass filter on low frequencies can be seen in Figure \ref{fig:RawvsGauss1}.

\section{Results}
\label{sec:Results}

To track the PM over all data sets, we started by determining which frequencies have a signal-to-noise ratio greater than four in all observations.
This is the case for the four strongest frequencies in the TESS data (the first four frequencies listed in Table \ref{tab:freqs}). 
The stability of those frequencies over the different observations is analyzed in Section \ref{sec:Frequency stability between the different observations}. 
We then looked at time delay curves created from simulated light curves. 
For this, the ``best-case scenario'' of a four-year Kepler observation of \bpic{} is studied in Section \ref{sec: bpic as seen by Kepler}. 
The time delays of the real observations and a comparison to a simulated re-creation can be found in Section \ref{sec: Observations and Simulations}. 
Finally, we analyze the pulsational stability of the \dsct{} pulsations of \bpic{} using TESS data in Section \ref{sec:stability}.

\subsection{Frequency stability between the different observations}
\label{sec:Frequency stability between the different observations}

\begin{figure}[ht!]
\centering
\includegraphics[width=0.5\textwidth]{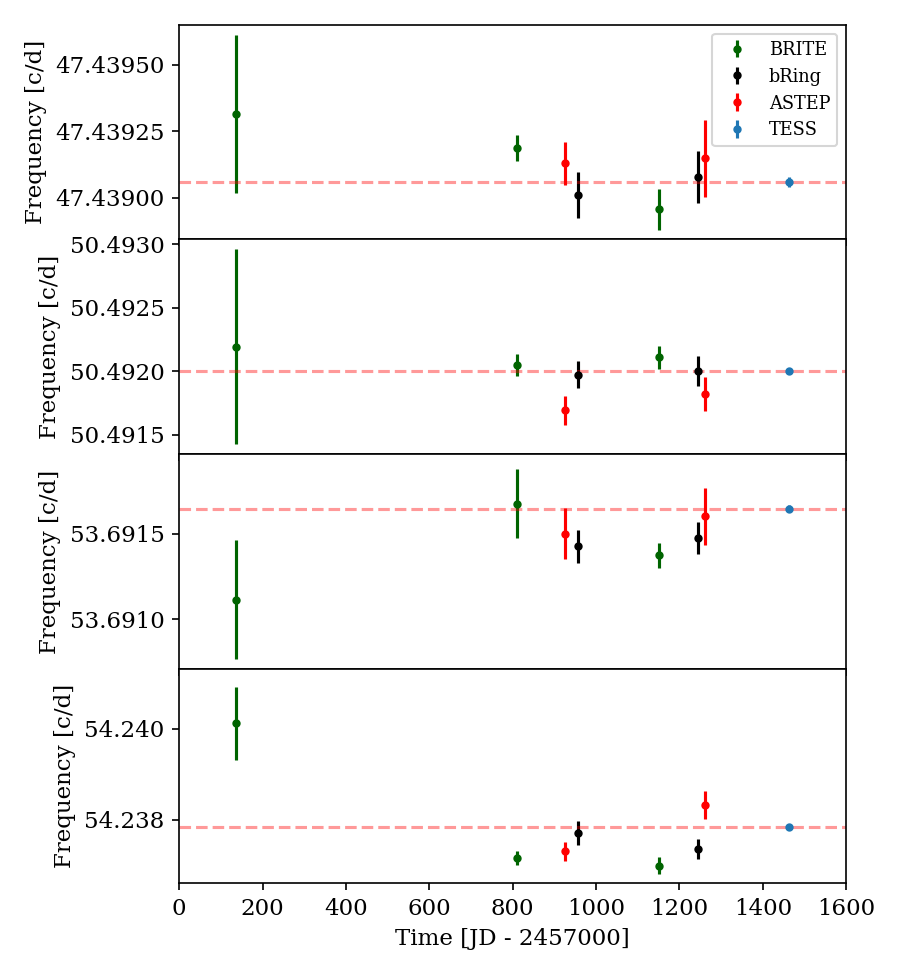}
\caption{Frequencies and their uncertainties over all data sets for the four modes that are visible in all observations from the four different observatories.
The dashed red line marks the frequency determined by the TESS mission, which has the smallest uncertainties. %
The uncertainties were calculated following \citet{Montgomery1999}.}
\label{fig:freqsmod}
\end{figure}

As mentioned in Section \ref{sec:The Phase Modulation Method}, the PM method derives a time delay from the observed PM at fixed frequencies. 
The precision with which pulsational frequencies can be determined depends on the quality of the data (cadence, timebase, precision, etc.). 
The photometry collected by the TESS mission has the smallest uncertainties in frequency of all data sets (see Fig. \ref{fig:freqsmod}). 
We therefore used TESS as a ``gold standard'' for the frequencies used in the PM method.
The uncertainties in the frequencies were calculated following \citet{Montgomery1999}. 
However, as noted in their publication, these errors are a lower limit of the true values. 
Keeping in mind that the actual error bars are probably bigger, one can see that the frequencies are in agreement with each other across the different data sets (Fig. \ref{fig:freqsmod}).

\subsection{Simulation based on Kepler data}
\label{sec: bpic as seen by Kepler}

\begin{figure}[ht!]
\centering
\includegraphics[width=0.5\textwidth]{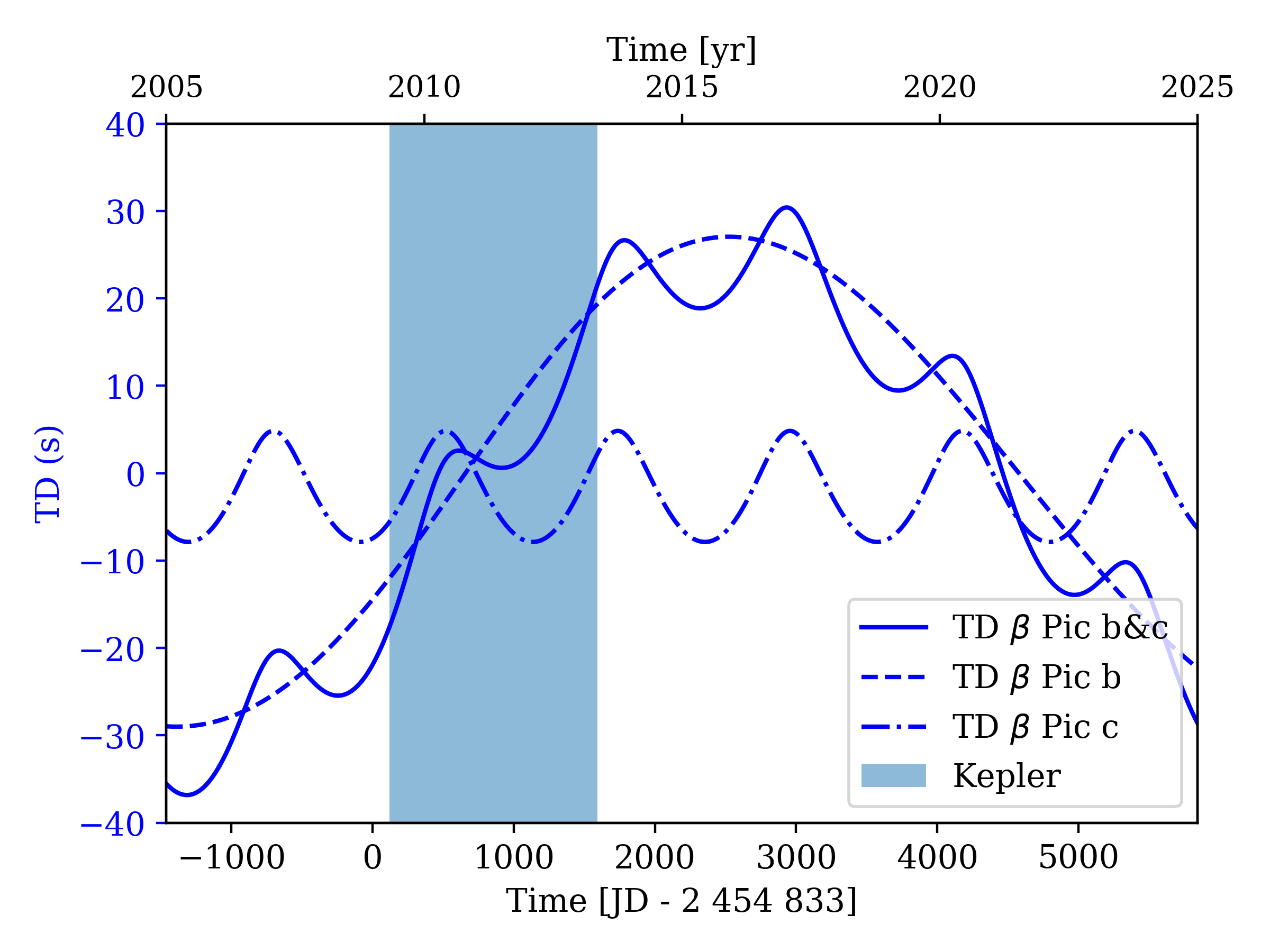}
\caption{Expected time delays for two planets in the \bpic{} system. The dashed line is for \bpicb{}, the dashed-dotted line is for \bpicc{}, and the solid line shows both.
The blue shaded region marks the time span of Kepler's four-year main mission; we note that Kepler did not observe \bpic{}.}
\label{fig:TD_Kepler}
\end{figure}

Figure \ref{fig:TD_Kepler} shows the expected time delays for \bpic{} caused by \bpicb{} and \bpicc{} using the full orbital solution given in Table \ref{tab:bpicb}. 
The addition of the individual contributions on the phases gives the total time delay curve as seen by the solid line in Figure \ref{fig:TD_Kepler}.

In order to see how such properties as photometric precision, cadence, and gaps in the observations influence the derived time delays, we simulated light curves of \bpic{}. 
We used the actual time stamps of the short-cadence observations of Kepler \citep{Borucki2010}, which have a cadence of around one minute.
The simulations consist of a multi-sine of the frequencies listed in Table \ref{tab:freqs}. 
Using even more frequencies increases the computational time without influencing the results of the simulations due to their low amplitudes. 
The time stamps were then modulated by the expected time delay at a given time using Equation \ref{eq:tau} and assuming a two-planet configuration in this system. 
Further, Gaussian noise on the order of 30 ppm was added to every data point, which is comparable to the noise floor of TESS.
Following the procedure explained in Section \ref{sec:The Phase Modulation Method}, the light curve was separated into 20-day segments, and the time delays were calculated from the phases in every segment with a fixed frequency. 
Finally, we calculated weighted time delay values and their corresponding uncertainties using the first three frequencies with the highest amplitudes.

\begin{figure}[ht!]
\centering
\includegraphics[width=0.5\textwidth]{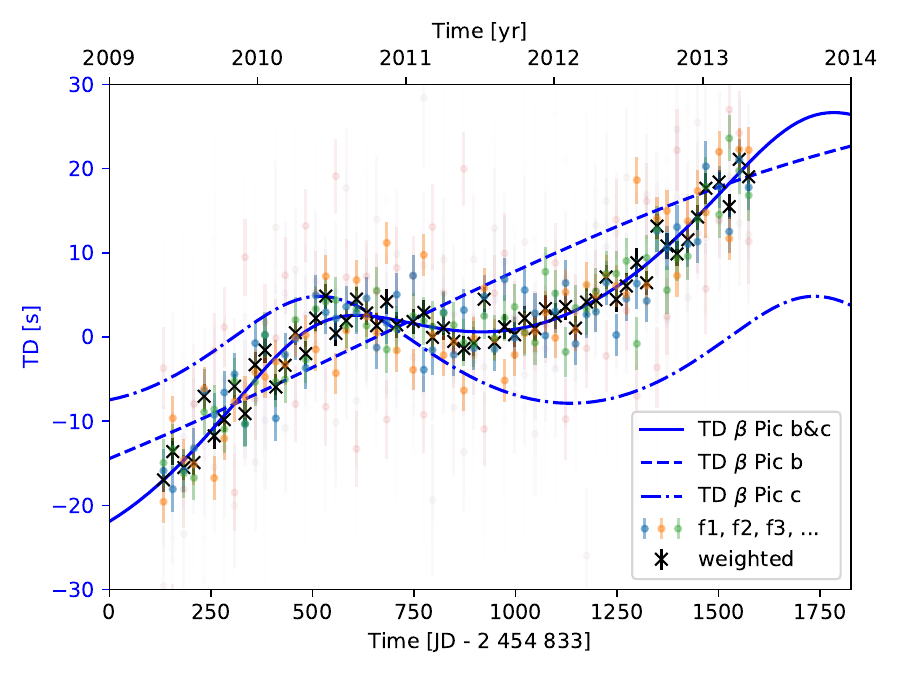}
\caption{Derived time delay curve using 20-day segments by simulating Kepler observations of \bpic.
Properties of the simulated light curve: one-minute cadence, continuous observations, 20 ppm noise in flux.
The strongest frequencies (f1, f2, f3, ...) are shown here with their uncertainties.
The weighted average of the measurements is shown in black.}
\label{fig:TD_Kepler1}
\end{figure}

\begin{figure}
\centering
\includegraphics[width=0.5\textwidth]{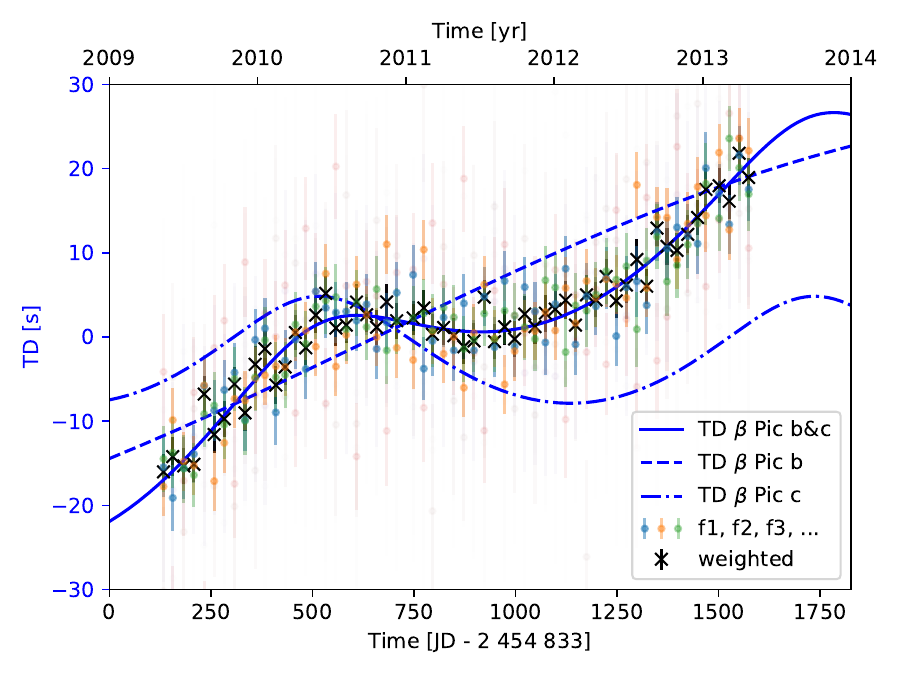}
\caption{Derived time delay curve using 20-day segments by simulating Kepler observations of \bpic.
Properties of the simulated light curve: two-minute cadence, continuous observations, 20 ppm noise in flux.}
\label{fig:TD_Kepler2}
\end{figure}

\begin{figure}
\centering
\includegraphics[width=0.5\textwidth]{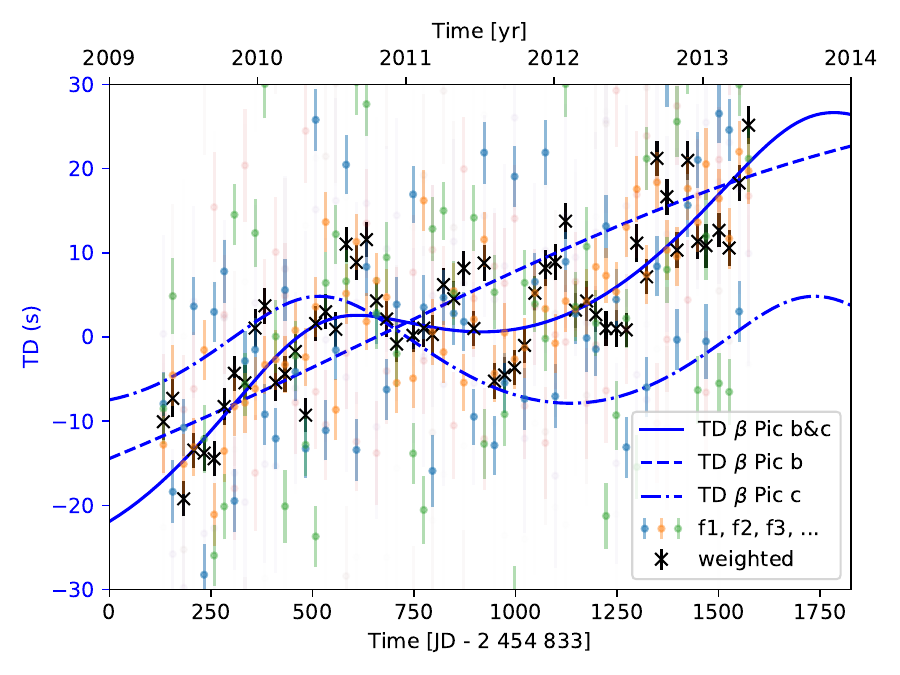}
\caption{{Derived time delay curve using 20-day segments by simulating Kepler observations of \bpic.
Properties of the simulated light curve: one-minute cadence, 0.5-day gaps every day, 20 ppm noise in flux.}}
\label{fig:TD_Kepler3}
\end{figure}

The first simulation (Fig. \ref{fig:TD_Kepler1}) used the Kepler short-cadence one-minute time stamps. 
The measured time delays follow the prediction for a two-planet case. 
Removing every second data point, which effectively reduces the number of measurements by 50\%, does not change the result (Fig. \ref{fig:TD_Kepler2}). 
One can, however, observe a small increase in the uncertainties for the time delays. 
This is expected, as a decrease of data points by a factor of two increases the uncertainty by a factor of $\sqrt{2}$, assuming Gaussian noise \citep{Murphy2012}. 
Finally, we induced gaps into the light curve, effectively simulating ground-based observations by having data only for half of the day. 
Due to a worse spectral window, there are many more peaks present in the amplitude spectrum. 
These new peaks influence the phases for the observed frequencies, as they cannot be resolved anymore. 
The scatter in the time delays increases significantly (Fig. \ref{fig:TD_Kepler3}), and one cannot conclusively distinguish between a one-planet solution (only \bpicb) or a two-planet solution.
A bigger segment size mitigates this effect, as expected by the Rayleigh criterion.

This clearly illustrates that gaps influence the time delays the strongest, as the uncertainties in phase only scale with the square root of the cadence factor. 
One should also remove identified frequencies that are not used in the time delay analysis to get rid of their aliases.

\subsection{Time delay analysis of the photometry}
\label{sec: Observations and Simulations}
\begin{figure}[ht!]
\centering
\includegraphics[width=1\columnwidth]{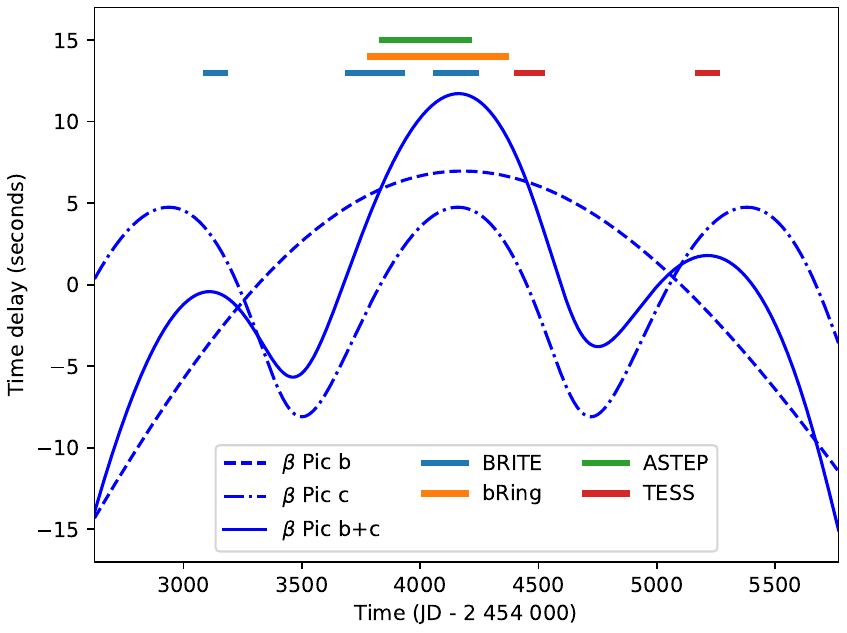}
\caption{Time delay predictions for \bpic{} b (dashed line), c (dashed-dotted line), and both planets (solid line). Times when the star was observed are marked with lines.}
\label{fig:TDRV_all_zoom}
\end{figure}

Figure \ref{fig:TDRV_all_zoom} shows the predicted time delays caused by the planets in the \bpic{} system during the times when the observatories BRITE, ASTEP, bRing, and TESS collected photometry for the star. 
The available observations have been introduced in Section \ref{sec:Observations}. 
The semi-amplitude of the predicted time delays for \bpicb{} and c is around 24 and 6 seconds, respectively.

\begin{figure}[ht!]
\centering
\includegraphics[width=0.5\textwidth]{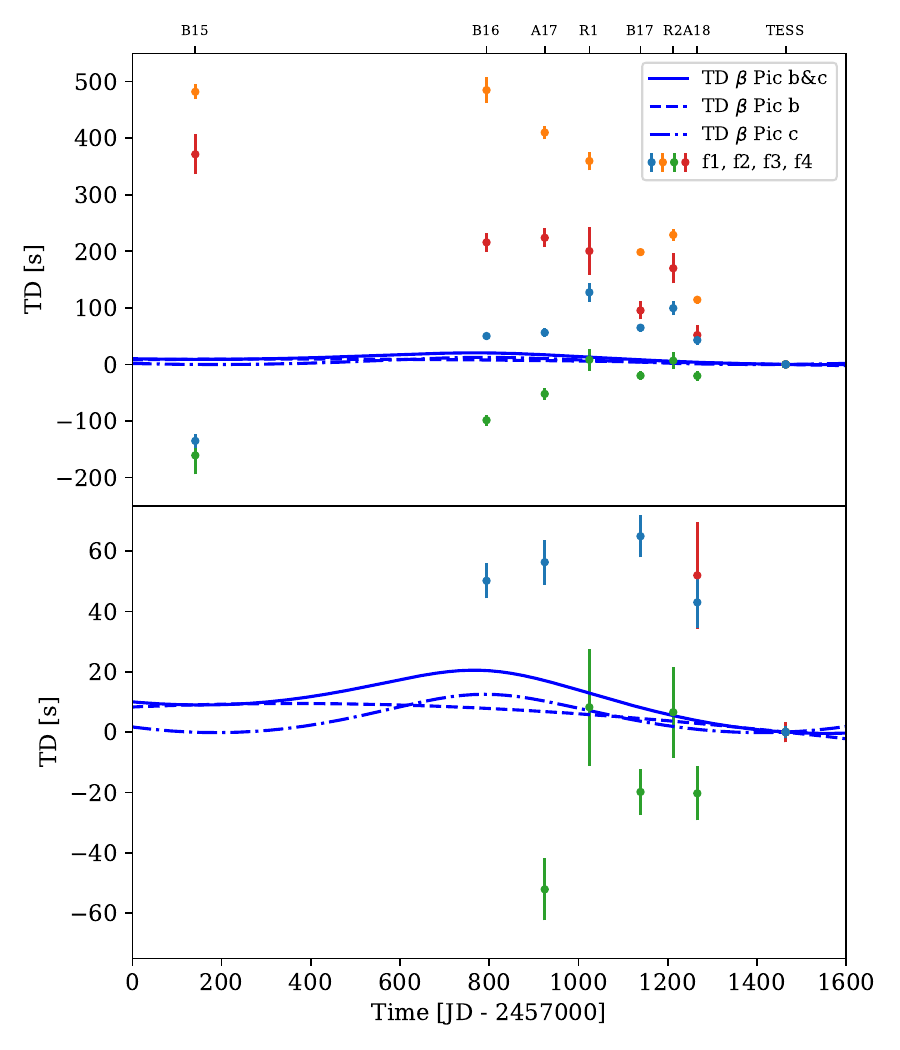}
\caption{Time delay plot calculated from the phases of four different frequencies for all available observations by BRITE, bRing, ASTEP and TESS. Each color represents a frequency (f1, f2, f3, and f4) listed in Table \ref{tab:freqs}. 
The blue lines indicate time delay predictions for \bpicb{} (dashed line), c (dashed, dotted line), and both planets (solid line).
The lower panel is a zoom-in of the upper panel.
The uncertainties in the time delays were derived from the covariance matrices given by the least-squares procedure, which was used in order to calculate the phases of the respective frequencies.
The ticks at the top of the plot denote the various observatories: B15, B16, and B17 for the BRITE observations in the years 2015, 2016, and 2017, respectively; A17 and A18 for the ASTEP observations in 2017 and 2018; and R1 and R2 for the first and second part of the bRing data.}
\label{fig:finall}
\end{figure}

\begin{figure}[ht!]
\centering
\includegraphics[width=0.5\textwidth]{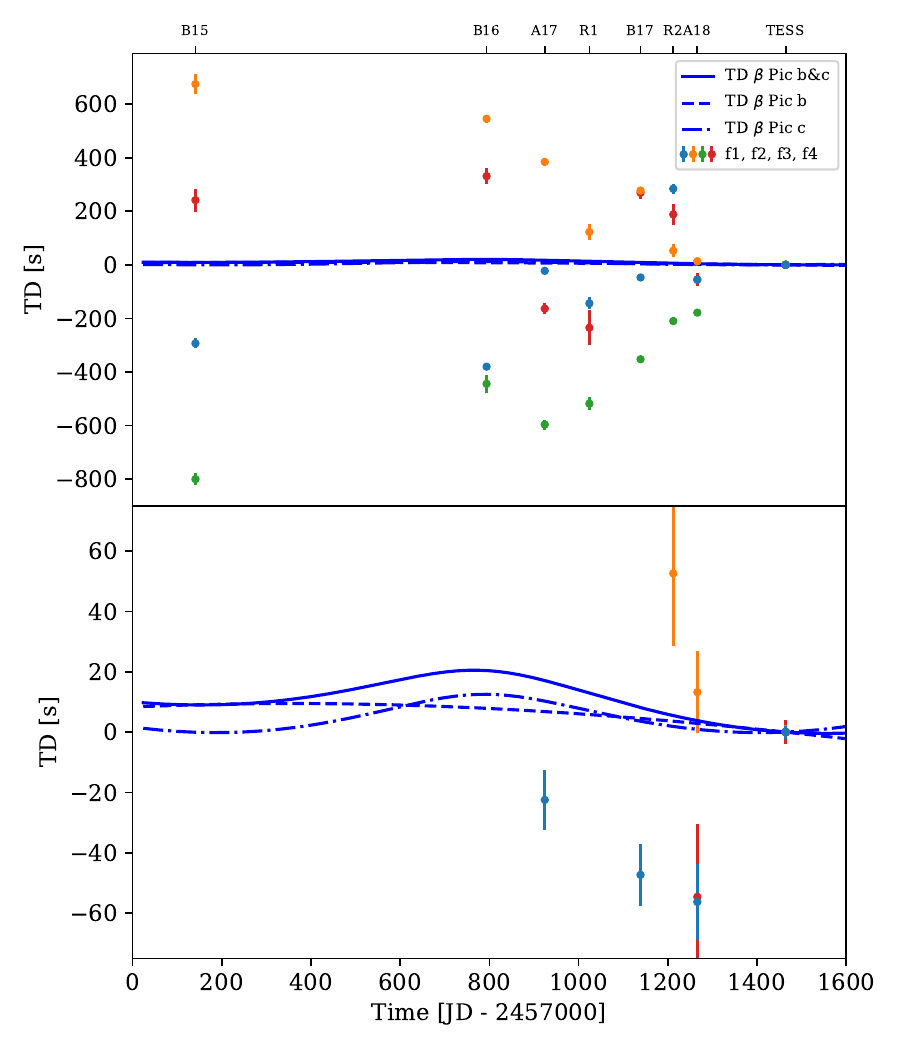}
\caption{Time delay plot for the simulated data set showing a high similarity to the time delays of the real data set presented in Figure \ref{fig:final2}.
The colored points represent the simulated time delays for the four strongest pulsational frequencies.
The blue lines indicate time delay predictions for \bpicb{} (dashed line), c (dashed-dotted line), and both planets (solid line).
The lower panel is a zoom-in of the upper panel.
A description of the ticks at the top of the plot can be found in the caption of Figure \ref{fig:finall}.}
\label{fig:final2}
\end{figure}

As seen in Section \ref{sec:Frequency stability between the different observations}, the TESS observations show the smallest uncertainties in frequency and were therefore used as a ``gold standard'' in this analysis. 
The frequency was thus fixed to the TESS values, as the PM method observes the phase shifts at a constant frequency (see Section \ref{sec:The Phase Modulation Method}). 
The time delay predictions (blue lines in Figure \ref{fig:TDRV_all_zoom}, \ref{fig:finall} and \ref{fig:final2}) were also normalized to the midpoint time of TESS. 
As a time delay is a relative measure and not an absolute one, we set the time delay for TESS to zero. 
The evaluated time delays shown in Figures \ref{fig:finall} and \ref{fig:final2} are therefore relative to the TESS values.

The code used to calculate the time delays was written for this analysis and is heavily based on existing ones, namely {\tt timedelay}\footnote{\url{https://github.com/danhey/timedelay}} and {\tt maelstrom}\footnote{\url{https://github.com/danhey/maelstrom}} \citep{Hey2020}. 
The equations that were needed in order to evaluate the time delays are given in Section \ref{sec:Time Delays} and \ref{sec:The Phase Modulation Method}.

The phases were calculated by subtracting the midpoint time of the full data set. 
As discussed in Section \ref{sec:Frequency stability between the different observations}, there are only four frequencies that are significant in all observations. 
The phases for each data set were then calculated based on a least-squares routine and their uncertainties from the respective covariance matrices. 
Equation \ref{eq:TD_ij} gives the conversion between the phase of a frequency and the respective time delays.

Figure \ref{fig:finall} shows the derived time delays for the four different frequencies. 
They are clearly not consistent with each other. 
As discussed in Section \ref{sec:4.7}, this rules out an extrinsic cause for the modulations (e.g., a companion), as all frequencies would show a similar behavior (examples of this are shown in Figure \ref{fig:TDsim}, \ref{fig:TD_Kepler1}, \ref{fig:TD_Kepler2} and \ref{fig:TD_Kepler3}). 
A change in frequency was ruled out in Section \ref{sec:Frequency stability between the different observations}. 
Furthermore,\citet{Zwintz2019} showed no significant phase change for our four frequencies in the BHr 2018 data set (therein, these four frequencies have the designations F8, F11, F13, and F15).\\

Next, we attempted to reproduce the different data sets as faithfully as possible and compare them with the time delay values shown in Figure \ref{fig:finall}. 
For that, we first determined the frequency, amplitude, and phase of the four pulsation modes visible in all observations. 
We calculated the residual noise for the pre-whitened data sets, which was then used to estimate the uncertainties following \citet{Montgomery1999}. 
As before, we fixed the frequency to the TESS value. 
As we do not know the exact ``true'' frequency of the pulsations with infinite precision, we introduced an offset between the true pulsational frequencies and the TESS data set in the simulations. 
The uncertainty in frequency for these four strongest frequencies is on the order of $10^{-5}$ \cd{} (see Table \ref{tab:freqs}). 
This offset explains the linear trend for every frequency that is visible in Figure \ref{fig:finall}. 
The time delays of the simulated data set are shown in Figure~\ref{fig:final2}.

This linear trend was discussed on simulated data in Section 3.2 of \citet{Murphy2016b} and ``almost certainly'' explains the observed trend in the WASP data of \citet{Murphy2013}. 
A way to correct for it is to evaluate the slope between two maxima or minima of the sinusoidal variations. 
This is not a possibility in our case, as we would have had to further segment the data sets to identify the position of the maxima or minima, leading to even higher scatter in the phase. 
The lower panel of Figure \ref{fig:finall} also shows that the uncertainties in the derived time delays for the data sets other than TESS are too big to differentiate between a one-planet or two-planet scenario, even without this linear trend. 
We therefore discuss the possibility of a second \bpic{} observation by TESS in the extended mission in the next section.

\subsection{Detection limits for \bpic{}}

\begin{figure}[ht!]
\centering
\includegraphics[width=1\columnwidth]{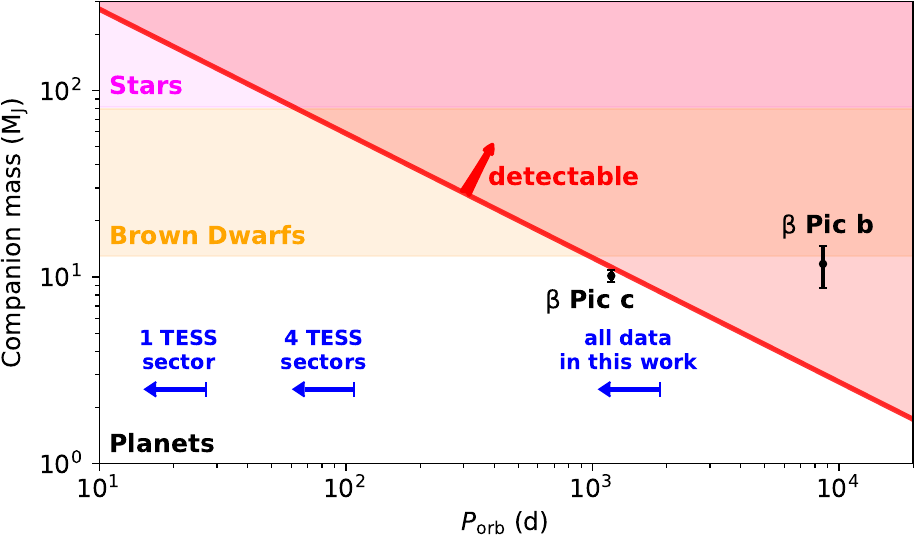}
\caption{
Detection limits for the \bpic{} system based on the calculations presented in \cite{Hey2020b} assuming 1.8 \Msun\ for the star and an inclination of $90^{\circ}$ for both planets. The detectable parameter space for companions depending on their orbital period $P_\textrm{orb}$ and mass is shaded red, assuming \bpic{} is a non-ideal \dsct{} pulsator (see Section \ref{sec:stability} for a discussion on the pulsational modes of \bpic{} as seen by TESS). \bpicc{} is generally not detectable due to the intrinsic noise of the pulsations, and the time delay caused by \bpicb{} has a period that is too long in comparison to the baseline of our observations. The masses corresponding to brown dwarfs (13 M$_\textrm{J}$ $\lesssim$ $M$ $\lesssim$ 80 M$_\textrm{J}$) are shaded in yellow, and the stellar regime ($M$ $>$ 80 M$_\textrm{J}$) is in magenta, with M$_\textrm{J}$ being a Jupiter mass. The figure is adapted from \cite{Hey2020b}.
}
\label{fig:detection_limits}
\end{figure}

Here, we determine the detection limits for companions in the \bpic{} system using the PM method. In a prior study by \cite{Hey2020b}, the authors tried to estimate the detection limits of companions around \dsct{} stars. Their approach involved simulating time-series observations of these pulsating stars and adding white noise to the data. This allowed them to find a relationship between the S/N of the stellar pulsations and the observed scatter in the resulting time delay ($a\,\sin i/c$). The established relationship could be directly converted into a parameter space defining detectable companion masses and their orbital periods. The authors found that the detectability of companions strongly depends on the S/N of the stellar pulsations.
Given the comparably low observed S/N of the \dsct{} pulsations observed in \bpic{} and the instability of the pulsational modes, we opted for a more conservative detection limit than what \citep[84$^{\textrm{th}}$ percentile in][]{Hey2020b} presented in their prior work. 

Our findings are presented in Figure \ref{fig:detection_limits}. We determined that the intrinsic variability of \bpic{} is too high to detect planet c. The other companion, \bpicb{}, is primarily not detectable at the moment due to the short baseline of observations compared to its long orbital period of approximately 24 years. It is worth noting that these calculations assume that the pulsational modes stay stable during the time of observations. However, we show in Section \ref{sec:stability} by using the TESS data that this assumption is generally not met, with some modes appearing and disappearing during the observations. This further complicates any efforts to detect companions around the star.

\subsection{Analysis of pulsational stability using TESS data}\label{sec:stability}

\begin{figure*}[ht!]
\centering
\includegraphics[width=0.975\textwidth]{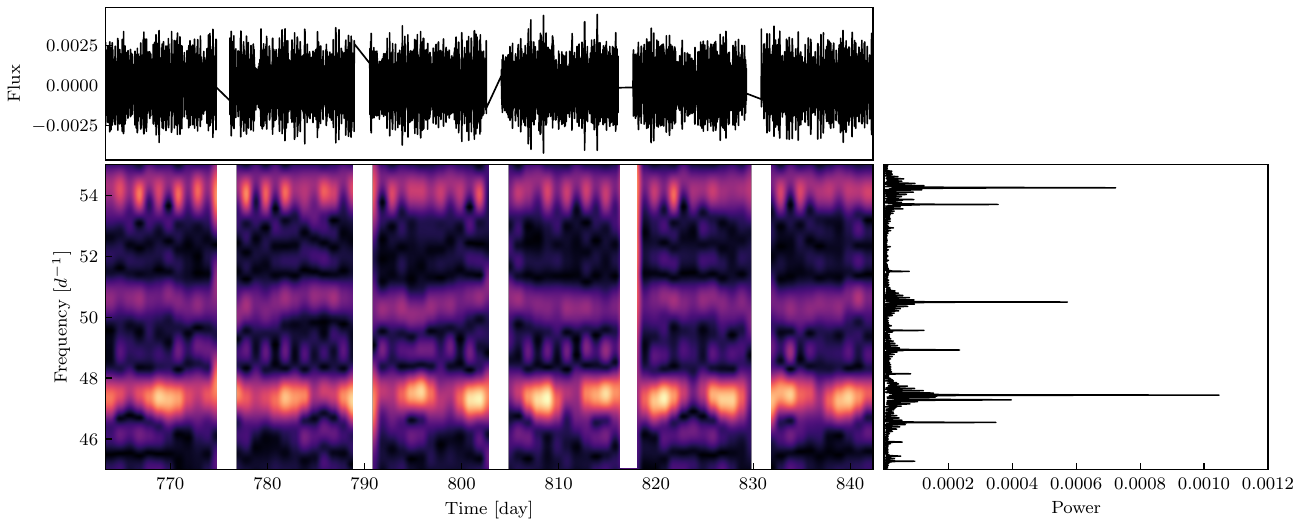}
\caption{Two-dimensional periodogram showing the frequency region between 45 and 55 d$^{-1}$.
One can see that most of the modes are significantly amplitude modulated by the rotational signal.
The mode at 54 d$^{-1}$ goes much faster, which is probably due to beating with nearby modes.
The mode at 50.5 d$^{-1}$ itself undergoes incoherent FM.}
\label{fig:hey1}
\end{figure*}

\begin{figure*}[ht!]
\centering
\includegraphics[width=0.975\textwidth]{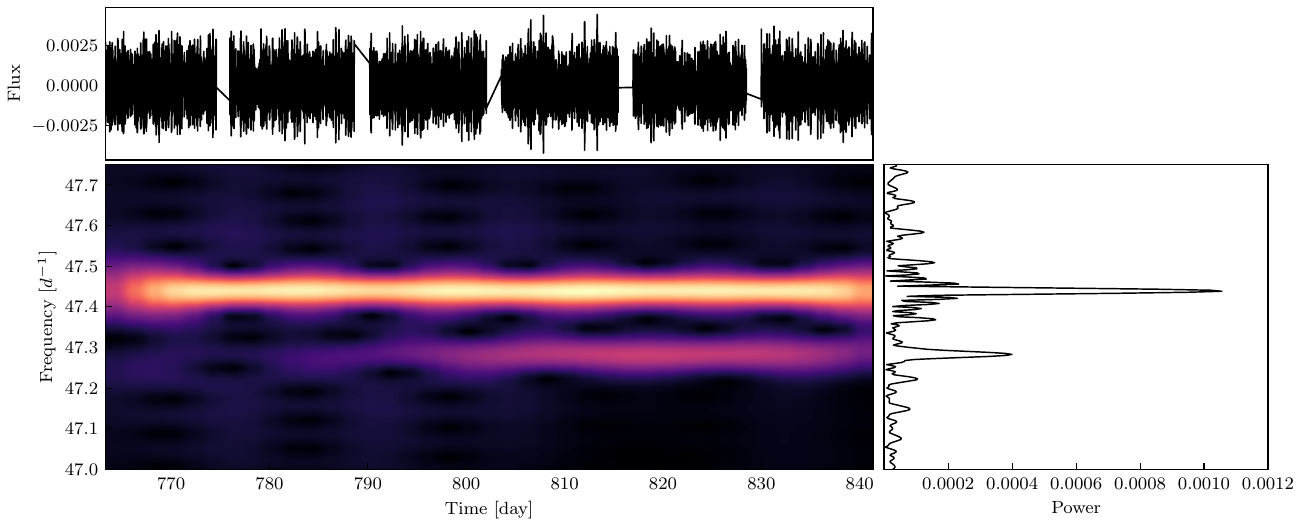}
\caption{Periodogram showing a pulsational mode around 47.3 d$^{-1}$ that is just appearing at the end of the observations.}
\label{fig:hey2}
\end{figure*}

\bpic{} was observed in seven individual sectors between October 2018 and February 2021 (see Table \ref{tab:obs}). 
We performed a frequency analysis of these TESS sectors using {\tt maelstrom}.
We find that $\beta$ Pictoris is seemingly undergoing significant frequency and amplitude modulation, which buries any signal induced by planetary companions.
The periodograms of the stellar pulsations clearly show significant amplitude modulation by the rotational signal (see Fig. \ref{fig:hey1}).
We also find that the star is showing modes that are appearing and disappearing on short timescales. 
Figure \ref{fig:hey2} shows a mode that seems to be just appearing during the second half of the TESS observations. 
In summary, we find that $\beta$ Pictoris' modes are not stable enough to probe to the necessary phase precision for the planetary companions.

%\newpage

\subsection{Comparison to KIC\,7917485}

Here we compare the \bpic{} system to another A star with a planet detected through pulsation timing, KIC\,7917485 \citep{Murphy2016a}, and evaluate the differences between the host stars that have affected the detectability of their corresponding planets. Compared to other Kepler \dsct{} stars, including those with binary star companions, \citep[e.g.,][]{Murphy2018a, Murphy2020b}, and \cite{Murphy2016a} found that KIC\,7917485 had an exceptionally low time delay noise (sixth lowest of the 2040 \dsct{} stars in the Kepler primary mission sample). Similar planetary-mass companions might exist around the second and ninth lowest noise stars in the Kepler sample (KIC\,9700322 and KIC\,8453431). Unlike KIC\,7917485, however, the observations of these systems and their potential companions do not cover a full orbital period, thus precluding determination of their orbital parameters.\\
The additional time-delay jitter seen in \bpic{} might arise from mode interaction or from other effects causing the observed changes in mode amplitude. It is this jitter that hinders the ability to detect planetary-mass objects around pulsating stars. This is comparable to the intrinsic RV jitter of a star, which impacts the ability to detect planets using RV measurements. The intrinsic variability is evident in the \dsct{} pulsations of \bpic{} and generally renders white dwarfs and subdwarfs as less ideal targets for the analysis of their time delays using the PM method \citep{Murphy2018b}. This limitation persists despite the high oscillation frequencies exhibited by these pulsating stars that would otherwise make them promising targets for detecting time delays caused by a companion \citep{Compton2016}.

\section{Conclusions}
\label{sec:Conclusions}

In this work, we have analyzed the time delays derived from the phases of the \dsct{} pulsations of \bpic. 
The photometric data of the star were collected over a time period of approximately four years by four different observatories: the BRITE-Constellation, bRing, ASTEP, and TESS. 
In contrast to previous studies, we did not segment the observations into smaller sets (e.g. ten-day bins). 
This would have caused high uncertainties in the phases and therefore also in the time delays.
Nevertheless, we could not see the influence of \bpicb{} or c in the data due to this time delay scatter. 
The uncertainty in the pulsational frequencies leads to a linear trend in the time delays and has also been seen in a previous study by \citet{Murphy2013} and in simulations by \citet{Murphy2016b}. %
We performed a frequency analysis using the open-source tool {\tt maelstrom}.
We find that $\beta$ Pictoris does not have the needed stability to detect planetary companions using the time delay method.
The stellar pulsations clearly show strong amplitude modulation caused by the rotational signal and identify modes that seemingly appeared during our observations.

Previous studies have used the PM method on Kepler data, finding many binary star systems \citep[][; and references therein]{Murphy2016b} and a planet \citep{Murphy2016a}. 
This work is the first to use the PM method with so many different data sets that have a precision significantly lower than the Kepler mission. 
However, the PM method remains valuable technique, as it is able to find planets and stars in a parameter space that is poorly covered by other methods, such as the RV method \citep[see e.g. ][]{Murphy2018b}. 

\begin{acknowledgements}

This work includes data collected by the TESS mission, which are publicly available from the Mikulski Archive for Space Telescopes (MAST). 
Funding for the TESS mission is provided by the NASA Explorer Program. 
This work has made use of data from the European Space Agency (ESA) mission {\it Gaia} (\url{https://www.cosmos.esa.int/gaia}), processed by the {\it Gaia} Data Processing and Analysis Consortium (DPAC, \url{https://www.cosmos.esa.int/web/gaia/dpac/consortium}).
Funding for the DPAC has been provided by national institutions, in particular the institutions participating in the {\it Gaia} Multilateral Agreement.
We made use of the software package {\tt Period04} \citep{Lenz2005}, the Python programming language \citep{Rossum1995}, and the open-source Python packages {\tt numpy} \citep{vanderWalt2011}, {\tt matplotlib} \citep{Hunter2007}, {\tt astropy} \citep{Astropy2013}, {\tt lightkurve} \citep{Lightkurve2018}, {\tt timedelay}\footnote{\url{https://github.com/danhey/timedelay}}, {\tt maelstrom}\footnote{\url{https://github.com/danhey/maelstrom}} and {\tt SMURFS} \citep{Mullner2020}.
This research has made use of the SIMBAD database, operated at CDS, Strasbourg, France.
This research has made use of the NASA Exoplanet Archive, which is operated by the California Institute of Technology, under contract with the National Aeronautics and Space Administration under the Exoplanet Exploration Program.
Part of this research was carried out at the Jet Propulsion Laboratory, California Institute of Technology, under a contract with the National Aeronautics and Space Administration (80NM0018D0004).
This research has made use of data collected by the BRITE-Constellation satellite mission, designed, built, launched, operated and supported by the Austrian Research Promotion Agency (FFG), the University of Vienna, the Technical University of Graz, the University of Innsbruck, the Canadian Space Agency (CSA), the University of Toronto Institute for Aerospace Studies (UTIAS), the Foundation for Polish Science \& Technology (FNiTP MNiSW), and National Science Centre (NCN).
The bRing observatory at Siding Springs, Australia was supported by a University of Rochester University Research Award.
The field activities at Dome C for ASTEP benefit from the support of the French and Italian polar agencies IPEV and PNRA in the framework of the Concordia station programme.
The PicSat team thanks funding from the European Research Council (ERC) under the Horizon 2020 research and innovation programme (Grant agreement No. 639248, LITHIUM).
SJM was supported by the Australian Research Council (ARC) through Future Fellowship FT210100485.
Finally, we thank the anonymous referee for a detailed report, which helped us
to improve the quality of this manuscript.

\end{acknowledgements}

\bibliographystyle{aa}
\bibliography{references}

\clearpage
\onecolumn 

\begin{appendix}

\section{TESS frequency analysis}

\subsection{Frequency list}

\begin{table*}[ht]
\centering
 \caption[]{Pulsational frequencies, amplitudes in instrumental millimagnitudes, and normalized flux in parts per million, phases, and signal-to-noise ratio sorted by the pre-whitening sequence.}\label{tab:freqs}
\renewcommand{\arraystretch}{1.35}
\begin{tabular}{l|l|l|l|l|l}
\hline\hline
\# & Freq. ($d^{-1}$) & Ampl. (mmag) & Ampl. (ppm) & Phase & S/N  \\\hline
1  & 47.43895(6)  & 1.029(9) & 948(9) & 0.9071(14) & 20.6 \\
2  & 53.69166(7)  & 0.948(9) & 873(9) & 0.2782(16) & 19.4 \\
3  & 50.49168(7)  & 0.926(9) & 852(9) & 0.5567(16) & 23.4 \\
4  & 54.23716(12) & 0.553(9) & 509(9) & 0.982(3)   & 22.8 \\
5  & 39.06315(15) & 0.442(9) & 407(9) & 0.699(3)   & 22.5 \\
6  & 46.54259(16) & 0.415(9) & 382(9) & 0.391(4)   & 18.7 \\
7  & 48.9192(3)   & 0.230(9) & 212(9) & 0.950(6)   & 17.3 \\
8  & 43.5283(3)   & 0.214(9) & 197(9) & 0.029(7)   & 19.9 \\
9  & 47.2853(4)   & 0.182(9) & 168(9) & 0.186(8)   & 16.2 \\
10 & 57.4525(4)   & 0.164(9) & 151(9) & 0.577(9)   & 18.0 \\
11 & 34.7605(5)   & 0.143(9) & 131(9) & 0.754(10)  & 23.7 \\
12 & 38.1297(5)   & 0.131(9) & 121(9) & 0.980(11)  & 20.2 \\
13 & 45.2698(5)   & 0.120(9) & 110(9) & 0.411(12)  & 12.8 \\
14 & 51.4969(6)   & 0.118(9) & 109(9) & 0.397(13)  & 14.7 \\
15 & 47.2686(7)   & 0.093(9) & 85(9)  & 0.419(16)  & 12.3 \\
16 & 50.8310(8)   & 0.086(9) & 79(9)  & 0.630(17)  & 12.9 \\
17 & 49.7131(8)   & 0.085(9) & 78(9)  & 0.290(17)  & 11.3 \\
18 & 53.8545(8)   & 0.085(9) & 78(9)  & 0.566(17)  & 9.5  \\
19 & 44.6833(8)   & 0.084(9) & 77(9)  & 0.297(18)  & 11.3 \\
20 & 65.1356(8)   & 0.083(9) & 76(9)  & 0.350(18)  & 17.9 \\
21 & 43.8292(8)   & 0.082(9) & 76(9)  & 0.555(18)  & 16.3 \\
22 & 49.5595(8)   & 0.079(9) & 73(9)  & 0.980(19)  & 13.4 \\
23 & 42.0365(9)   & 0.077(9) & 71(9)  & 0.327(19)  & 11.1 \\
24 & 54.2269(9)   & 0.073(9) & 67(9)  & 0.51(2)    & 10.0 \\
25 & 41.6498(9)   & 0.071(9) & 65(9)  & 0.59(2)    & 13.0 \\
26 & 48.1381(10)  & 0.064(9) & 59(9)  & 0.23(2)    & 11.2 \\
27 & 45.8998(10)  & 0.064(9) & 59(9)  & 0.73(2)    & 12.4 \\
28 & 50.2689(12)  & 0.054(9) & 50(9)  & 0.90(3)    & 12.1 \\
29 & 75.6780(13)  & 0.052(9) & 48(9)  & 0.68(3)    & 12.2 \\
30 & 58.3469(13)  & 0.050(9) & 46(9)  & 0.61(3)    & 11.3 \\
31 & 45.4375(14)  & 0.047(9) & 44(9)  & 0.00(3)    & 12.3 \\
32 & 54.4625(14)  & 0.047(9) & 43(9)  & 0.16(3)    & 8.1  \\
33 & 53.6827(15)  & 0.042(9) & 39(9)  & 0.16(3)    & 7.3  \\
34 & 53.5521(16)  & 0.040(9) & 37(9)  & 0.48(4)    & 7.7  \\
35 & 42.1735(16)  & 0.040(9) & 37(9)  & 0.70(4)    & 9.5  \\
36 & 58.2515(17)  & 0.039(9) & 36(9)  & 0.81(4)    & 10.5 \\
37 & 42.3963(17)  & 0.039(9) & 36(9)  & 0.68(4)    & 11.5 \\\hline
\end{tabular}
%\tablefoot{\\
%}

\end{table*}

\clearpage

\subsection{Gaussian high-pass filter}
\label{sec:Gaussian Highpass filter}

\begin{figure}[h!]
\centering
\includegraphics[width=0.5\textwidth]{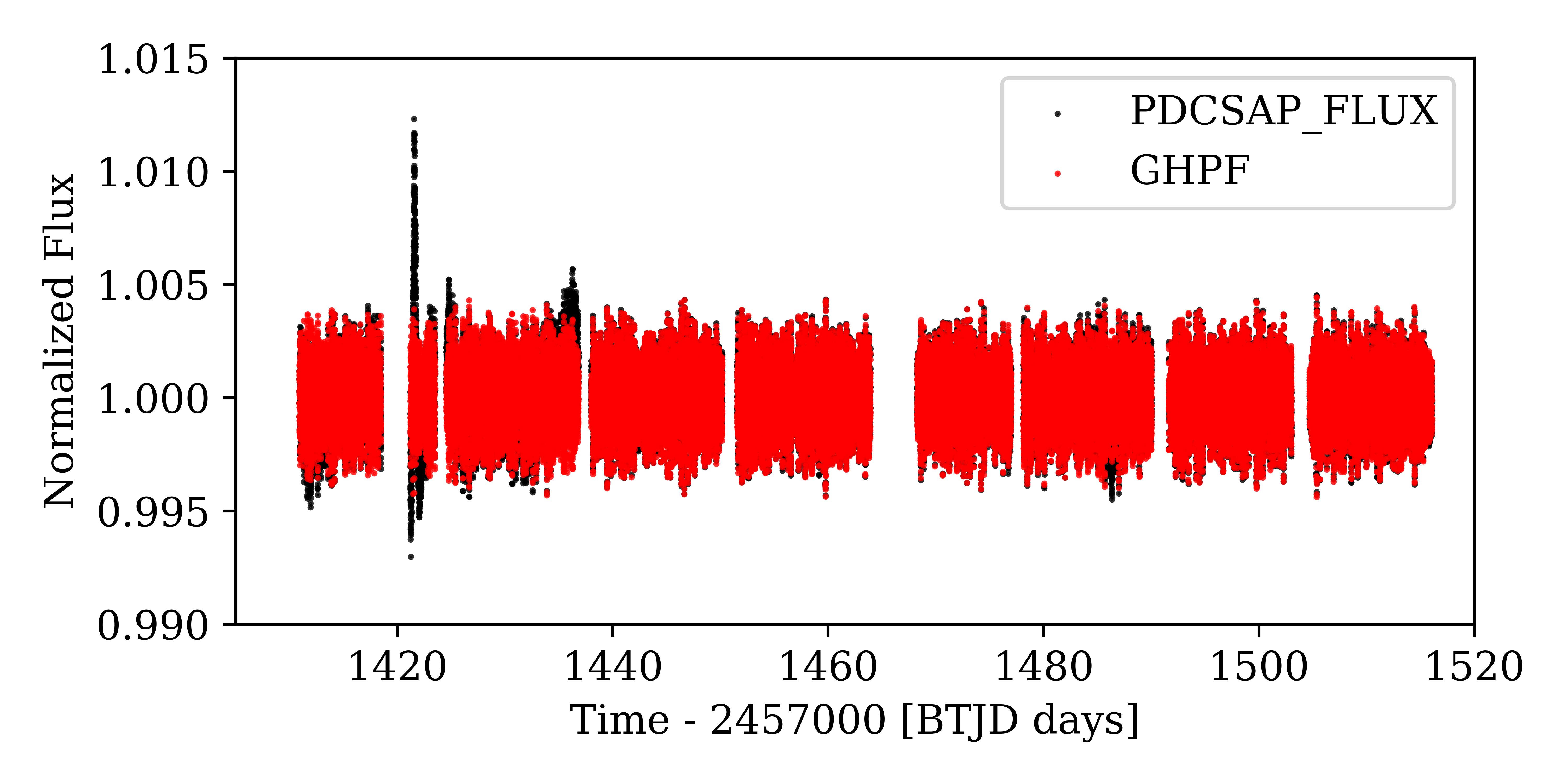}
\caption{Comparison of the PDCSAP light curve (black in the background) and the Gaussian high-pass filter (red dots) of it. The Gaussian high-pass filter clearly shows less long-term variations in the light curve.}
\label{fig:gaussian_lc_comp}
\end{figure}

By applying a Gaussian high-pass filter on the TESS light curve (see Fig. \ref{fig:gaussian_lc_comp}), the long-term variations such as systematics and the exocomets are significantly weakened. At the same time however, the higher frequencies related to the \dsct{} pulsations are preserved. 

\end{appendix}

\end{document}